\documentclass[%
 reprint,
superscriptaddress,
nofootinbib,
 amsmath,amssymb,
 aps,
floatfix,
]{revtex4-1}

\newcommand{\grafe}[1]{\left\{ #1 \right\}}
\newcommand{\tonde}[1]{\left( #1 \right)}
\newcommand{\quadre}[1]{\left[ #1 \right]}

\usepackage{xcolor}
\usepackage{graphicx}
\usepackage{dcolumn}
\usepackage{amsthm, amsfonts}
\usepackage{bbm}
\usepackage{comment}
\usepackage{caption}
\usepackage{subcaption}
\usepackage[normalem]{ulem}
\usepackage{empheq}

\usepackage{bm}

\newcommand{\VR}[1]{\textrm{\textcolor{blue}{#1}}}

\begin{document}

\preprint{APS/123-QED}
\title{Curvature-driven pathways interpolating between stationary points:\\ the case of the pure spherical $3$-spin model }

\author{Alessandro Pacco}
\email{alessandro.pacco@universite-paris-saclay.fr}
\affiliation{%
 Université Paris-Saclay, CNRS, LPTMS, 91405 Orsay, France
}%
\author{Giulio Biroli}
\affiliation{
Laboratoire de Physique de l’Ecole Normale Supérieure, ENS, Université PSL, CNRS,
Sorbonne Université, Université de Paris, F-75005 Paris, France
}%

\author{Valentina Ros}%
\affiliation{%
 Université Paris-Saclay, CNRS, LPTMS, 91405 Orsay, France
}%

\date{\today}

\begin{abstract}
This paper focuses on characterizing the energy profile along pathways connecting different regions of configuration space in the context of a prototypical glass model, the pure spherical $p$-spin model with $p=3$. The study investigates pairs of stationary points (local minima or rank-1 saddles), analyzing the energy profile along geodesic paths and comparing them with ``perturbed" pathways correlated to the landscape curvature. The goal is to assess the extent to which information from the local Hessian matrices around stationary points can identify paths with lower energy barriers. Surprisingly, unlike findings in other systems, the direction of softest local curvature is not a reliable predictor of low-energy paths, except in the case in which the direction of softest curvature corresponds to an isolated mode of the Hessian. However, other information encoded in the local Hessian does allow the identification of pathways associated with lower energy barriers.  We conclude commenting on implications for the system's activated dynamics. 
\end{abstract}

\maketitle
\onecolumngrid
Understanding the statistical properties of random energy or cost landscapes in high dimensions is a rather ubiquitous problem. In several cases one is interested in characterizing the profile of the landscape along pathways connecting different regions of configuration space; these regions may correspond to different local minima of an energy function \cite{wales1998archetypal,jonsson1998nudged} (equivalently, to fitness local maxima \cite{mauri2022mutational,tian2020exploring}), or to different configurations reached by algorithms optimizing a cost function \cite{draxler2018essentially,freeman2016topology,annesi2023star}. 
In many situations, in particular in several glassy systems~\cite{mezard1987spin,berthier2011theoretical,parisi2020theory}, the height of the typical effective barriers crossed during the dynamics increases at low temperatures, thus leading to super-Arrhenius behaviors \cite{debenedetti2001supercooled}. Characterizing the paths crossing these barriers and connecting low-energy  configurations is a major challenge to develop the theory of glassy dynamics. Note that there are also other kinds of complex high-dimensional landscapes: some with 
a preferred minimum, as in the problem of protein folding~\cite{onuchic1997theory}, or with many flat minima separated by low barriers, as in machine learning~\cite{semerjian2008freezing, draxler2018essentially, garipov2018loss, annesi2023star}.  
 Here, we focus on the first class of energy landscapes and address the challenge outlined above. \\
We consider a prototypical glass model, the spherical $p$-spin model with $p=3$,
 given by a random Gaussian function defined on a high-dimensional sphere~\cite{crisanti1992sphericalp,crisanti1993spherical}. This landscape 
  exhibits a multiplicity of stationary points (local minima, but also saddles) that are isolated in configuration space, and separated by energy barriers.
 We consider pairs of such stationary points and characterize the energy profile along simple paths interpolating between them. In particular, we compute the typical energy profile along the geodesic path, determining how the resulting energy barrier depends
on the energy of the stationary points and on their distance in configuration space. We then compare the geodesic energy profile with the profile along “perturbed"  pathways, which follow directions in configuration space that are correlated to the landscape curvature around one of the two stationary points.\\
Our goal is to investigate to what extent one can identify good pathways (i.e., paths associated with lower energy barriers) using the information on the landscape around the two stationary points, encoded in their local Hessian. This question is motivated by studies of finite-dimensional systems of jammed and mildly supercooled particles. For the former, the softest Hessian modes at a given configuration provide directions toward low energy barriers for particle rearrangements \cite{xu2010anharmonic}. For the latter, the spatial amplitudes of the softest Hessian modes correlate with the regions where irreversible rearrangements of particles take place \cite{widmer2008irreversible}.  Contrary to these findings, we show that in the case of the purely random landscape we consider in this work, the direction of softest local curvature is not a predictor of paths with lower energy barriers, except in the case in which the local Hessian has an isolated mode. However, we argue that having access to the whole local Hessian in general allows to identify pathways associated to lower energy barriers. In the conclusion we also discuss possible physical reasons for the discrepancy between our findings and the ones cited above.  \\

\begin{figure}[ht]
\centering
\includegraphics[width=0.7\textwidth]{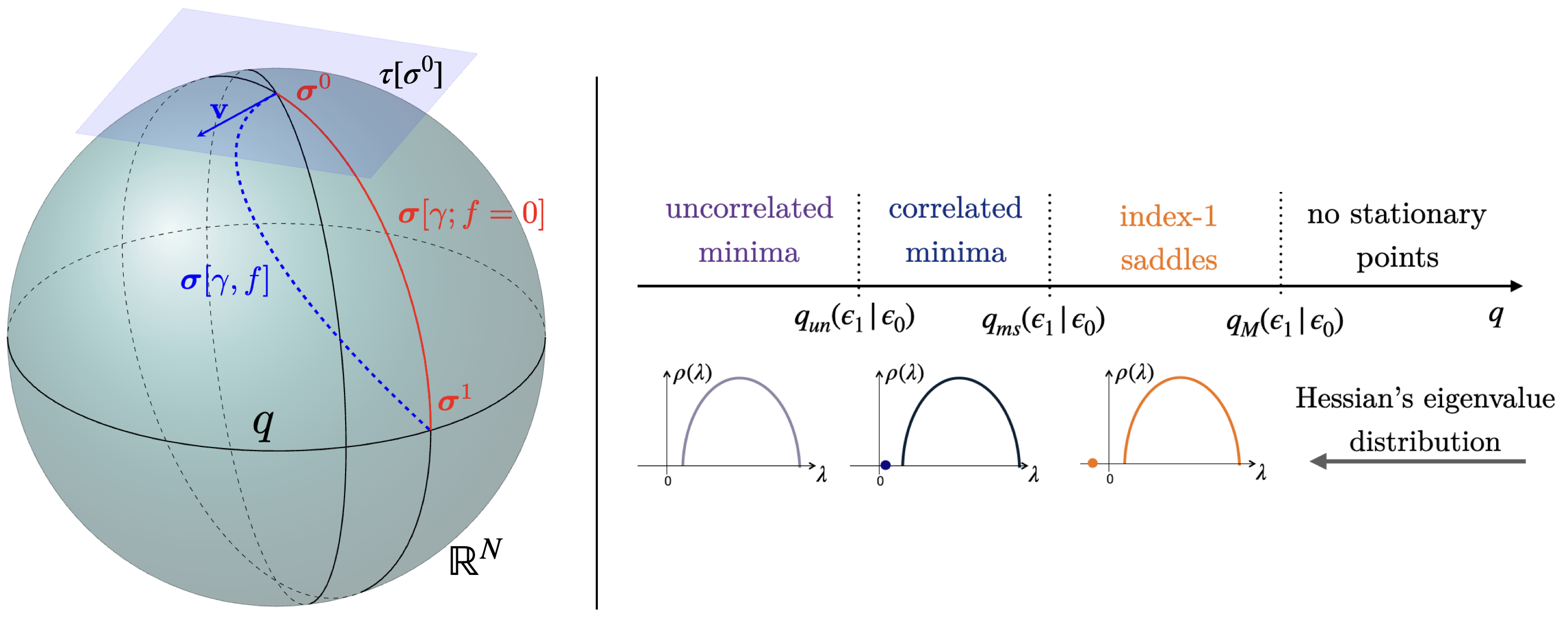}\\
\vspace{.5 cm}
\caption{ {\it Left.} Sketch of the configuration space (sphere), with a geodesic path (red thick line) and a perturbed path (blue dashed line) interpolating between two configurations ${\bm \sigma}^a$ with $a=0,1$.  The tangent plane $\tau[{\bm \sigma}^0]$ to which ${\bf v}$ belongs is also sketched. {\it Right.} Eigenvalue distribution of the Hessian at ${\bm \sigma}^1$ as a function of the overlap $q$ between the ${\bm \sigma}^a$, for fixed $\epsilon_a$.}
\label{fig:SketchPath}
\end{figure}

{\it The structure of the energy landscape. }The energy function of the pure $p$-spin spherical model reads: 
\begin{equation}
\label{eq:landscape}
\mathcal{E}({\bf s}) = \sqrt{\frac{ p!}{2 N^{p-1}}}\sum_{i_1<\ldots <i_p}a_{i_1\ldots i_p} s_{i_1}\ldots  s_{i_p},
\end{equation}
where ${\mathbf{s}} =( s_1,\ldots,  s_N)\in\mathbb{R}^N$ 
satisfies ${\mathbf{s}}^2=\sum_{i=1}^N  s_i^2=N$.
The couplings $a_{i_1\ldots i_p}$ are i.i.d Gaussian variables with zero mean and unit variance: we denote with $\mathbb{E}\left[ \cdot  \right]$  the average with respect to them. The energy \eqref{eq:landscape} is a Gaussian function with covariance
\begin{equation}\label{eq:Enp}
\mathbb{E}\left[ \mathcal{E}({\bf s}^0)  \mathcal{E}({\bf s}^1)\right] = \frac{N}{2} \,  \quadre{q({\bf s}^0, {\bf s}^1)}^p,
\end{equation}
where $q({\bf s}^0, {\bf s}^1)= N^{-1} \tonde{{\bf s}^0\cdot {\bf s}^1}$ is the overlap between the configurations, which measures how close they are in configuration space. 
We denote the energy density of a configuration by $\epsilon({\bf s}) = \lim_{N\to\infty}N^{-1}\,\mathcal{E}({\bf s})$. There are two peculiar values of energy density for this model, $\epsilon_{\rm gs}$ and $\epsilon_{\rm th}$~\cite{crisanti1992sphericalp, cavagna1998stationary, cavagna1997investigation, cavagna1997structure, cavagna1999quenched, auffinger2013random, subag2017complexity}. For $\epsilon<\epsilon_{\rm gs}$ no configuration is present for typical realizations of the randomness: $\epsilon_{\rm gs}$ is the typical energy of the deepest minima, i.e. the ground states. For  $\epsilon>\epsilon_{\rm gs} $ the landscape presents a number of stationary points (minima, saddles, maxima) that scales exponentially with $N$.
Their stability changes at the threshold  $\epsilon_{\rm th}=-\sqrt{2 (p-1)/p}$: below $\epsilon_{\rm th}$ the exponential majority of the stationary points are local minima; the number of saddles with intensive index $k=o(N)$ \cite{auffinger2013random, auffinger2020number} (the index counting the number of directions in configuration space along which the curvature of the landscape is negative) is also exponentially large below $\epsilon_{\rm th}$, even though smaller than that of minima. Above $\epsilon_{\rm th}$ the stationary points are overwhelmingly saddles with $k=\mathcal{O}(N)$. The arrangement of these points 
 in configuration space has also been investigated \cite{cavagna1997investigation, cavagna1998stationary,ros2019complexity, subag2017complexity,ros2020distribution}, by determining the typical number  $\mathcal{N}_{{\bf s}^0}(\epsilon,q |\epsilon_0)$ of stationary points ${\bf s}^1$ that are at fixed energy density $\epsilon_1\in \quadre{\epsilon_{\rm gs}, \epsilon_{\rm th}}$ and overlap ${\bf s}^0\cdot{\bf s}^1= N q$ with a reference minimum ${\bf s}^0$ of the landscape at arbitrary energy density $\epsilon_0 \in \quadre{\epsilon_{\rm gs}, \epsilon_{\rm th}}$. For large $N$, the scaling of the typical value of $\mathcal{N}_{{\bf s}^0}(\epsilon,q |\epsilon_0)$   is controlled by the
constrained complexity~\cite{ros2019complexity,ros2020distribution}:
\begin{align}
\label{eq:quenchedcomp}
\Sigma(\epsilon_1,q|\epsilon_0)=\lim_{N\to\infty}\frac{\mathbb{E}\left[\log\mathcal{N}_{{\bf s}^0}(\epsilon_1,q |\epsilon_0)\right]}{N},
\end{align}
where the average is over both the population of local minima at energy density $\epsilon_0$ at fixed randomness, and over the randomness. From now on, we focus on parameters for which $\Sigma(\epsilon_1,q|\epsilon_0) \geq 0$: negative values of this function indicate the region of parameters where stationary points are atypical, i.e., they are found with a probability that decays exponentially fast with $N$. By monitoring the stability of the stationary points at fixed $\epsilon_a$ (with $a=0,1$), one finds several transitions as a function of $q$, represented schematically in Fig.~\ref{fig:SketchPath}. For $q$ smaller than a threshold $q_{un}(\epsilon_1|\epsilon_0)$ they are typically local minima with an Hessian spectrum that shows no correlations with ${\bf s}^0$. For intermediate $q_{un}(\epsilon_1|\epsilon_0) \leq q \leq q_{ms}(\epsilon_1|\epsilon_0)$ they are minima that are correlated to ${\bf s}^0$, in the sense that the eigenvector associated to the minimal eigenvalue of their Hessian (which identifies the direction of minimal curvature of the landscape around the  minimum) is oriented towards ${\bf s }^0$.
For large $q$, the stationary points are typically saddles of index $k=1$: the landscape has a single direction of negative curvature, which is oriented towards the reference minimum ${\bf s}^0$. For $q$ larger than a threshold $q_{M}(\epsilon_1|\epsilon_0)$, no stationary point is present. See Appendix \ref{app:recap_paper_saddles} for a concise account of these results. In this energy landscape thus there are exponentially many (in $N$) saddles with energy below $\epsilon_{\rm th}$ which surround the reference minimum, from which ${\bf s }^0$ is reachable by following the direction of negative curvature of the saddle \cite{ros2021dynamical}. These saddles are possible escape states for the system when it is dynamically trapped in ${\bf s}^0$, and they are likely to be the first energy barriers that the system crosses in its dynamics, even though one expects frequent returns to ${\bf s}^0$ after these barriers crossings \cite{ros2021dynamical}. 
Given two local minima at arbitrary overlap $q$ with each others, however, it is unknown how high in energy  the system has to climb in order to transition between them. In particular, it is unknown if this energy barrier is always much higher than $\epsilon_{\rm th}$, or whether one can find pairs of distant minima that are connected by a sequence of saddles having all energies below $\epsilon_{\rm th}$. To get a proxy of these barriers, we compute the energy profile along simple paths interpolating between pairs of stationary points at arbitrary overlap $q$. \\

{\it Interpolating paths and energy profiles.}  We introduce the vectors ${\bm\sigma}={\bf s}/\sqrt{N}$ belonging to the $N$-dimensional sphere of unit radius, $\mathcal{S}_N(1)$, and the scaled energy function $h({\bm\sigma}):=\mathcal{E}({\bm\sigma}\sqrt{N})\sqrt{2/N}$. We denote with ${\bf g}({\bm\sigma})$ and ${\bf \mathcal{H}}({\bm\sigma})$ the gradient and the Hessian matrix of the function $h({\bm\sigma})$ restricted to $\mathcal{S}_N(1)$. For a fixed realization of the landscape, we consider two configurations ${\bm \sigma}^0$  and ${\bm \sigma}^1$ drawn at random from the population of stationary points of $h({\bm \sigma})$ such that ${\bf g}({\bm \sigma}^a)=0$ for $a=0,1$. We extract the stationary points in such a way that each ${\bm \sigma}^a$ has energy density $\epsilon_a \in [\epsilon_{\rm gs}, \epsilon_{\rm th}]$ for $a=0,1$, and their overlap ${\bm \sigma}^0 \cdot {\bm \sigma}^1$ equals to some  $q\in[0,1]$. We are interested in the energy density profile along paths lying on the surface of $\mathcal{S}_N(1)$, which interpolate between the two stationary points. We parametrize the paths as follows:
\begin{align}
\label{eq:path}
    {\bm\sigma}[\gamma;f]=\gamma{\bm \sigma}^1 + \beta[\gamma; f]\; {\bm \sigma}^0 + f(\gamma){\bf v},\quad \gamma\in[0,1],
\end{align}
where $f$ is a continuous function such that $f(0)=f(1)=0$, and $\bf v$ is a norm-1 vector orthogonal to both ${\bm \sigma}^0$ and ${\bm \sigma}^1$. 
The condition that ${\bm \sigma}[\gamma;f]$ lies on $\mathcal{S}_N(1)$ enforces
\begin{equation}
\label{eq:beta}
\beta[\gamma; f] = - \gamma q+ \sqrt{1- \gamma^2(1-q^2)- [f(\gamma)]^2}.
\end{equation}
Notice that this quantity has to be real, and this imposes some constraints on $f(\gamma)$. 
When $f \equiv 0$, Eq.~\eqref{eq:path} gives the geodesic path connecting the two stationary points. The function  $f$ acts as a perturbation of the geodesic, along the direction identified by the vector ${\bf v}$, see Fig.~\ref{fig:SketchPath}. Therefore, we are restricting to interpolating paths belonging to the low-dimensional subspace spanned by the vectors ${\bf v}$ and $ {\bm \sigma}^a$, intersected with the surface of the sphere. 
We aim at computing the typical energy profile 
\begin{equation}\label{eq:Profile}
\epsilon_{\bf v}[\gamma;f]:= \lim_{N \to \infty}  \mathbb{E}\quadre{\frac{h({\bm\sigma}[\gamma;f])}{\sqrt{2N}}},
\end{equation}
where the average is over the distribution of stationary points ${\bm \sigma}^0, {\bm \sigma}^1$ with energy densities $\epsilon_0$, $\epsilon_1$ and overlap $q$, and over the realizations of the landscape. The average over ${\bm \sigma}^0$ and $ {\bm \sigma}^1$ requires some care, since it can be performed within two different protocols (referred to as \emph{quenched} and \emph{annealed} in the terminology of glasses \cite{franz1998effective}); in Appendix~\ref{app:setup} we discuss both averaging schemes, arguing that for the problem at hand they give identical results. \\
To specify our choice of ${\bf v}$ in \eqref{eq:path}, it is convenient to introduce the vectors  
\begin{align}
\label{eq:def_basis}
    {\bf e}_{N-1}({\bm \sigma}^0)= \frac{q \; {\bm \sigma}^0-{\bm \sigma}^1}{\sqrt{1-q^2}},\quad
    {\bf e}_{N-1}({\bm \sigma}^1)= \frac{q \; {\bm \sigma}^1-{\bm \sigma}^0}{\sqrt{1-q^2}}.
\end{align}
Each $ {\bf e}_{N-1}({\bm \sigma}^a)$ belongs to the tangent plane $\tau[{\bm \sigma}^a]$ to $\mathcal{S}_N(1)$ at the point ${\bm \sigma}^a$ (meaning that ${\bf e}_{N-1}({\bm \sigma}^a) \perp {\bm \sigma}^a$), and identifies the direction in the tangent plane pointing towards the other configuration ${\bm \sigma}^{b \neq a}$. Let also ${\bf x}_i$ with $i=1, \cdots, N-2$ be an arbitrary basis of the subspace orthogonal to the ${\bm \sigma}^a$. Finally, we denote with $\lambda_{\rm min}^a$ the minimal eigenvalue of $\mathcal{H}({\bm \sigma}^a)$ and with  
${\bf e}_{\rm min}({\bm \sigma}^a)$ the associated eigenvector, and let: 
\begin{equation}
    u^a=({\bf e}_{\rm min}({\bm \sigma}^a)\cdot {\bf e}_{N-1}({\bm \sigma}^a))^2.
\end{equation}
We consider two possible choices for ${\bf v}$. The first corresponds to
\begin{equation}
\label{eq:v_start}
{\bf v}\to{\bf v}_{\rm soft}^a= \frac{{\bf e}_{\rm min}({\bm \sigma}^a)- \sqrt{u^a}\,  {\bf e}_{N-1}({\bm \sigma}^a)}{\sqrt{1-u^a}},
\end{equation}
meaning that the path is deformed in the directions of softest curvature of the energy landscape at ${\bm \sigma}^a$. The second one corresponds to 
\begin{equation}
\label{eq:v_start2}
{\bf v}\to{\bf v}_{\rm Hess}= Z \sum_{i=1}^{N-2} [{\bf x}_i \cdot \tilde{\mathcal{H}}({\bm \sigma}^0) \cdot {\bf e}_{N-1}({\bm \sigma}^0)] \, {\bf x}_i
\end{equation}
where  $\tilde{\mathcal{H}}({\bm \sigma}^a)= \mathcal{H}({\bm \sigma}^a)+ 3\sqrt{2N} \, \epsilon_a$ denote the Hessian matrices at the stationary points shifted by a constant, and $Z$ a normalization constant. Consider now ${\bf v} \to {\bf v}_{\rm Hess}$. This choice is motivated by the study of the gradient vector ${\bf g}$ at each configuration ${\bm \sigma}(\gamma)$ along the geodesic path. As we show in Appendix~\ref{app:gradient_max_barr}, at each point the gradient has a tangent component ${\bf g}^\parallel$ to the path, and an orthogonal component ${\bf g}^\perp$ that is proportional to \eqref{eq:v_start2}. 
While ${\bf g}^\parallel$ obviously vanishes at the value of $\gamma$ that corresponds to the local maximum of the geodesic energy profile, ${\bf g}^\perp$ does not, meaning that the maximum of the geodesic profile is not a stationary point of $h({\bm \sigma})$. 
This suggests that interpolating paths associated to lower barriers can be found by deforming the geodesic path in the direction of ${\bf g}^\perp$, since this is likely the direction that the path would follow if it was allowed to relax in configuration space by gradient descent \cite{jonsson1998nudged,bolhuis2002transition}. In fact, Ref. \cite{freeman2016topology} developed an iterative algorithm to find low energy paths for deep neural networks based on this idea.\\

\emph{The case $p=3$ and the underlying random matrix problem.} From now on we focus specifically on $p=3$. This restriction is motivated by the fact that in this case the energy profile \eqref{eq:Profile} can be expressed as a function of the local properties of the landscape at ${\bm \sigma}^a$ only, i.e. of the local gradients ${\bf g}({\bm \sigma}^a)$ and Hessian matrices $\mathcal{H}({\bm \sigma}^a)$. By implementing the constraints $h({\bm \sigma}^a)= \sqrt{2 N} \epsilon_a$, ${\bf g}({\bm \sigma}^a)=0$ and by using the fact that typically $h({\bf v})=0$, one obtains:
\begin{equation}\label{eq:PrePAth}
\begin{split}
\epsilon_{\bf v}[\gamma,f]&=
\tonde{\gamma^3 +3\gamma^2 \beta \,q} \epsilon_1+ \tonde{\beta^3+3 \beta^2 \gamma \,q} \epsilon_0 - \sqrt{1-q^2} \gamma \beta f  \; \mathbb{E}\quadre{{\bf v}\cdot \frac{ \tilde{\mathcal{H}}({\bm \sigma}^0) }{\sqrt{2N}}\cdot {\bf e}_{N-1}({\bm \sigma}^0)}\\
& +   \frac{f^2}{2} \mathbb{E}\quadre{  \gamma \, {\bf v} \cdot \frac{ \tilde{\mathcal{H}}({\bm \sigma}^1) }{\sqrt{2N}}\cdot {\bf v} + \beta\, {\bf v} \cdot \frac{ \tilde{\mathcal{H}}({\bm \sigma}^0) }{\sqrt{2N}}\cdot {\bf v}},
\end{split}
\end{equation}
where we omitted the dependence of $\beta$ on $\gamma$ and $f$. It follows that for $p=3$ and with our choices of ${\bf v}$,  the energy profile depends only on correlations between the entries of the Hessian matrices, whose statistics has to be conditioned to the properties of ${\bm \sigma}^a$. This statistics is described in detail in Appendix~\ref{app:statistics_HESSIANS}. With the choice \eqref{eq:v_start}, in particular, the profile depends on the matrix elements of the Hessian  $\tilde{\mathcal{H}}({\bm \sigma}^1)$ on the minimal eigenvector of the Hessian ${\mathcal{H}}({\bm \sigma}^0)$ (equivalently, $\tilde{\mathcal{H}}({\bm \sigma}^0)$), or vice-versa. For any $q>0$, the two matrices are correlated (due to the fact that the landscape at the two points is correlated as \eqref{eq:Enp}) and thus the matrix element is non-trivial. As we argue in Appendix \ref{app:Overlaps}, to determine its typical behavior for large $N$ one needs to compute the typical value of the overlap between arbitrary eigenvectors of the two correlated matrices. This random matrix problem is discussed in recent  literature for standard random matrix ensembles \cite{bun2016rotational,bun2018overlaps}; for the particular type of random matrix ensembles describing the statistics of the Hessians $\mathcal{H}({\bm \sigma}^a)$, this overlap function has been computed explicitly in \cite{paccoros}, and the results are recalled in Appendix \ref{app:Overlaps}. Below, we discuss the properties of the energy profiles that are obtained making use of these results. \\

\begin{figure}[ht]
\centering
\includegraphics[width=0.58\textwidth]{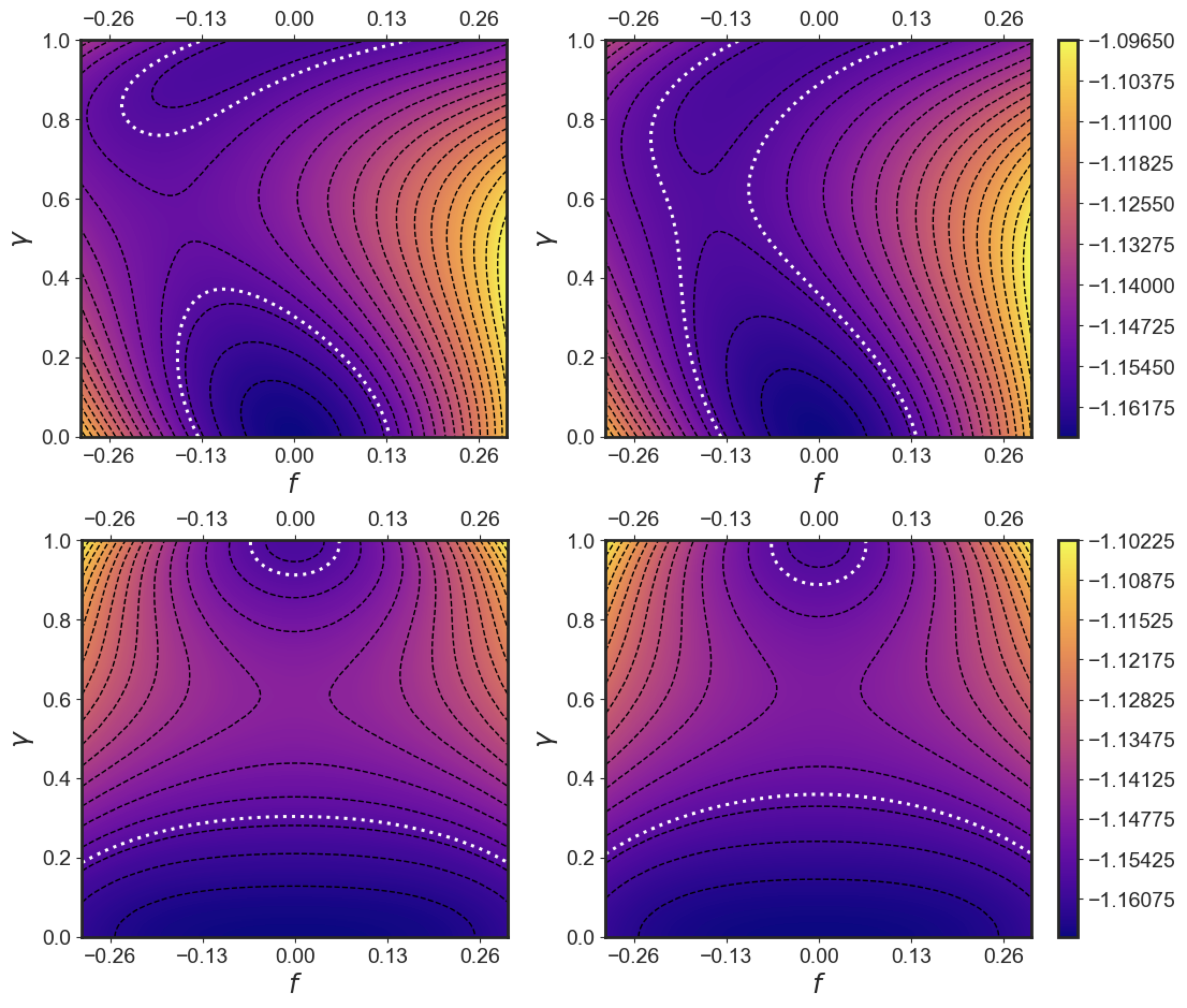}
\caption{Color plot of $\epsilon_{\bf v_{\rm soft}}[\gamma;f]$ as a function of $\gamma$ and $f$ for $\epsilon_0=-1.167$,  $\epsilon_1=-1.157$ and  $q=0.66 \in [q_{un},q_{ms}]$ (\textit{left}) and $q=0.7>q_{ms}$ (\textit{right}).  The plots correspond to ${\bf v}_{\rm soft} \to {\bf v}_{\rm soft}^1$ (\textit{top}), and  ${\bf v}_{\rm soft} \to {\bf v}_{\rm soft}^0$ (\textit{bottom}).}
\label{fig:density_plots}
\end{figure}

\emph{Results.} 
For ${\bf v}\to{\bf v}_{\rm soft}^0$, the interpolating path is as in  Fig.~\ref{fig:SketchPath}. When $\epsilon_0 < \epsilon_{\rm th}$, the reference stationary point ${\bm \sigma}^0$ is typically a local minimum: the shifted Hessian $N^{-\frac{1}{2}}{\tilde{\mathcal{H}}({\bm \sigma}^0)}$ has the statistics of matrices extracted from a Gaussian Orthogonal Ensemble (GOE) with variance $6/N$. The eigenvalues distribution of the shifted Hessian follows a semicircular law, and the minimal eigenvalue equals to  $-2\sqrt{6}$. The associated eigenvector will show no condensation in the direction of ${\bf e}_{N-1}({\bm \sigma}^0)$, meaning that  typically $u^0$ takes the value $u^0_{\rm typ}=0$. The path along \eqref{eq:path} then reads (see Appendix \ref{app:resulting_energy_profiles} for a derivation)
\begin{align}
\label{eq:energy_profile_quench_0}
\begin{split}
&\epsilon_{{{\bf v}_{\rm soft}^0}}[\gamma;f]=\tonde{\gamma^3 +3\gamma^2 \beta[\gamma;f] \,q}  \epsilon_1+  (\beta^3[\gamma;f]+3 \beta^2[\gamma;f] \gamma \,q)\epsilon_0-\sqrt{3}\,f^2(\gamma)\left( \beta[\gamma;f]+\gamma\,q\right).
\end{split}
\end{align}
One finds that $\delta \epsilon_{{{\bf v}_{\rm soft}^0}}/\delta f=0$ is satisfied by $f \equiv 0$, and that for $\epsilon_a <\epsilon_{\rm th}$ it holds $\delta^2 \epsilon_{{{\bf v}_{\rm soft}^0}}/\delta f^2>0$ at $f \equiv 0$, meaning that the geodesic path is a minimum of the functional~\eqref{eq:energy_profile_quench_0}.
Arbitrary  deformations of the interpolating path in the direction of softest curvature at ${\bm \sigma}^0$ go through regions of the landscape of higher energy density, on average. This observation is confirmed by Fig.~\ref{fig:density_plots} \textit{bottom}, which shows two density plots of Eq.~\eqref{eq:energy_profile_quench_0} as a function of $\gamma$ and $f$, where $f$ is allowed to take any value within its range of validity that keeps $\beta$ in \eqref{eq:beta} well defined. Arbitrary paths are obtained drawing curves connecting the points $f(0)=0$ and $f(1)=0$ in a continuous and {injective} fashion. One finds that the energy profile along these curves is non-monotonic, with a local maximum whose energy we refer to as the energy barrier. In Fig.~\ref{fig:density_plots} the parameters $\epsilon_0$ and $\epsilon_1$ are fixed, while $q$ is tuned in such a way that ${\bm \sigma}^1$ is either a minimum with an Hessian with a single isolated eigenvalue (\textit{left}) or an index-1 saddle (\textit{right}). The white dotted lines represent the level curves of value $\epsilon_\text{th}$. 
The plot confirms that the lowest energy barrier is obtained for the geodesic path $f\equiv 0$, and shows that such barrier is well above $\epsilon_{\rm th}$.
The same results are obtained for different values of $\epsilon_a$ and $q$. One finds that the barrier associated to the geodesic path increases when the overlap $q$ decreases and when $\epsilon_1$ increases towards $\epsilon_{\rm th}$.
 We remark that the profile \eqref{eq:energy_profile_quench_0} is not the same that one would obtain choosing ${\bf v}$ in \eqref{eq:PrePAth} as a purely random Gaussian vector ${\bf v}_{\rm rand}$, uncorrelated to the local Hessian: with that choice, all the terms depending on ${\bf v}$ in Eq. \eqref{eq:PrePAth} vanish on average, and one obtains 
$\epsilon_{{{\bf v}_{\rm rand}}}[\gamma;f]=\tonde{\gamma^3 +3\gamma^2 \beta\,q}  \epsilon_1+  (\beta^3 +3 \beta^2\gamma \,q)\epsilon_0.$
For fixed $f(\gamma)$ this energy profile is systematically higher than \eqref{eq:energy_profile_quench_0}. However, the functional is again minimized by $f \equiv 0$.

The case ${\bf v}\to{\bf v}_{\rm soft}^1$ is richer. In this case, depending on the values of $\epsilon_a$ and $q$, the stationary point ${\bm \sigma}^1$ is either a rank-1 saddle ($q_{sm} < q \leq q_M$), a minimum with one isolated mode in the Hessian ($q_{un} < q  \leq q_{sm}$), or an uncorrelated minimum ($q<q_{un}$), see Fig. \ref{fig:SketchPath}. Whenever the Hessian at ${\bm \sigma}^1$ has an isolated eigenvalue ($q>q_{un}$), it is the smallest eigenvalue and its typical value and the typical value $u^1_{\rm typ}$ are given by
\begin{equation}\label{eq:Typicals}
\begin{split}
&\lambda_{\rm typ}^1=
\frac{3 \, \delta \epsilon_q}{\sqrt{2}q }
\quadre{\frac{1+3 q^2}{
   1-q^2}
      - \sqrt{1-  \frac{2(1-q^2)^2}{3(1+q^2)\delta \epsilon_q^2}}}\quad,\quad\quad u_{\rm typ}^1=\frac{1+q^2}{1+3q^2 -\frac{\sqrt{2} (q-q^3) \lambda_{\rm typ}^1 }{3 \, \delta \epsilon_q}}\quadre{1- \frac{(1-q^2) \, g( \lambda_{\rm typ}^1)}{12 \sqrt{2} (q+q^3) \delta \epsilon_q} }
      \end{split}
\end{equation}
where $\delta \epsilon_q=\epsilon_0-q \epsilon_1$ and $g(\lambda)=1+3 q^2 \lambda- \text{sign}(\lambda) (1-q^2)\sqrt{\lambda^2- 24}$, see Appendix~\ref{app:statistics_HESSIANS}. In this case the energy profile reads: 
\begin{align}
\label{eq:energy_quench_1}
\begin{split}
\epsilon_{{{\bf v}_{\rm soft}^1}}[\gamma;f]&=(\gamma^3+3\gamma^2\beta\,q)\epsilon_1+
(\beta^3+3\gamma\beta^2\,q)\epsilon_0 + \gamma\,\beta\, f
\sqrt{\frac{u^1_{\rm typ}(1-q^2)}{2(1-u^1_{\rm typ})}}
\left(\frac{6 \sqrt{2} \, q \, (\epsilon_0-q \epsilon_1)}{1-q^2}-\lambda_{\rm typ}^1\right)\\
&+\frac{f^2\,(\gamma + 2 \beta)\,  u^1_{\rm typ} }{2\sqrt{2}(1-u^1_{\rm typ})}\left(\frac{6 \sqrt{2} \, q \, (\epsilon_0-q \epsilon_1)}{1-q^2}-\lambda_{\rm typ}^1\right)+\frac{f^2 \,\beta\,u^1_{\rm typ} }{2\sqrt{2}(1-u^1_{\rm typ})}\frac{6 \sqrt{2} q (q \epsilon_0-\epsilon_1)}{1-q^2} +\frac{f^2\,\gamma}{2\sqrt{2}}\lambda_{\rm typ}^1\\
& +\frac{f^2 \,\beta\, }{2\sqrt{2}(1-u^1_{\rm typ})}\int_{-2\sqrt{6}}^{2\sqrt{6}} d\lambda \,\frac{\sqrt{36 - \lambda^2}}{18 \, \pi } \,\Phi(\lambda_{\rm typ}^1,\lambda)\,\lambda.
\end{split}
\end{align}
 The function $\Phi$ in the last term of this expression gives the typical value of the overlap between the eigenvector associated to $\lambda^1_{\rm min}$ and any arbitrary eigenvector of $\tilde{H}({\bm \sigma}^0)$ with eigenvalue $\lambda$. Its expression is rather involved, and we report it in Appendix~\ref{app:Overlaps} (See Eq.~\eqref{eq:app:Phi_iso_bulk}). 
\begin{figure}[ht]
\centering
\includegraphics[width=0.51\textwidth]{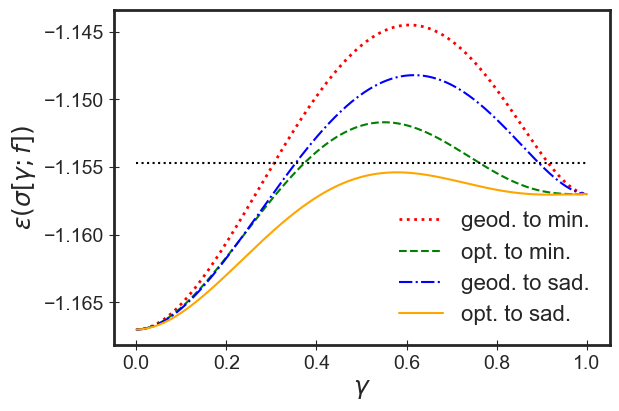}
\caption{Comparison between the energy profile along the geodesic (geod.) and optimal (opt.) paths, for the same parameters as in Fig.~\ref{fig:density_plots}({\it top}) which correspond to ${\bm \sigma}^1$ being a minimum (min.) and index-1 saddle (sad.).}
\label{fig:barriers}
\end{figure} 
 
In Fig.~\ref{fig:density_plots} (\textit{top}) we show two density plots associated to \eqref{eq:energy_quench_1}: clearly in this case the geodesic path is no longer optimal, and thus the energy barrier is lowered by deforming the path in the direction of softest curvature at ${\bm \sigma}^1$. The optimal path is obtained numerically, by selecting the lowest energy point for each $\gamma$ and by verifying that this leads to a continuous path. In Fig.~\ref{fig:barriers} we compare the energy profile \eqref{eq:energy_quench_1} evaluated along the geodesic and optimal paths, for  ${\bm \sigma}^1$ being a correlated minimum  ($q_{un} < q \leq q_{ms}$) or a saddle ($q_{ms} < q < q_M$). In the latter case, the optimal path lies entirely below $\epsilon_{\rm th}$ (dashed horizontal line). When ${\bm \sigma}^1$ is an uncorrelated minimum, since $u_{\rm typ}^1=0$ the behavior is analogous as in Eq.~\eqref{eq:energy_profile_quench_0} with $\gamma$ and $\beta$ exchanged in the last term of the expression; the optimal path is again the geodesic one.  \\

\begin{figure}[ht]
\centering
\includegraphics[width=0.5\textwidth]{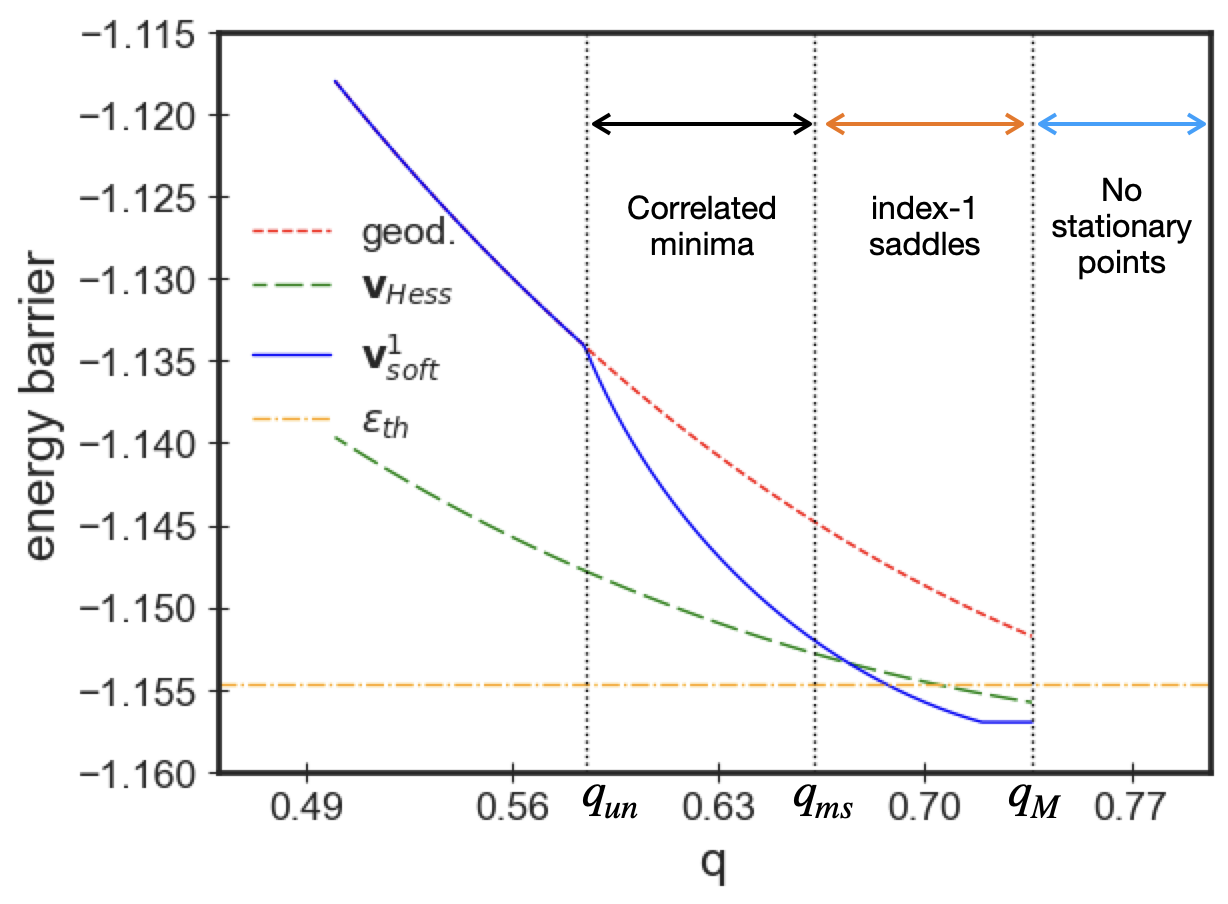}
\caption{Energy barrier along the geodesic and optimal (perturbed) paths as a function of $q$, for $\epsilon_0=-1.167$, $\epsilon_1=-1.157$ and ${\bf v}$ being equal to either $ {\bf v}^1_{\rm soft}$ or ${\bf v}_\text{Hess}$.}
\label{fig:energy_barrier}
\end{figure}

Similarly as above, plugging ${\bf v}_{\rm Hess}$ into \eqref{eq:PrePAth} and computing the averages one finds (see Appendix \ref{app:gradient_max_barr}):
\begin{align}
\label{eq:energy_profile_Hess}
\begin{split}
&\epsilon_{{{\bf v}_{\rm Hess}}}[\gamma;f]=\tonde{\gamma^3 +3\gamma^2 \beta[\gamma;f] \,q}  \epsilon_1+  (\beta^3[\gamma;f]+3 \beta^2[\gamma;f] \gamma \,q)\epsilon_0- \gamma \beta[\gamma;f] f(\gamma)\;\sqrt{\frac{3(1-q^2)^2}{1+q^2}}.
\end{split}
\end{align}
In this case the barrier along the perturbed path is lower than the geodesic one even for $q<q_{un}$, even though it is still above the threshold for all choices of $\epsilon_a < \epsilon_{\rm th}$.
In Fig.~\ref{fig:energy_barrier} we plot a comparison between the energy barrier of the geodesic path and that of the optimal paths, for given $\epsilon_a$ and varying $q$. The barriers decrease with $q$; when ${\bf v}={\bf v}_{\rm Hess}$ the deformed path is always associated to a lower energy barrier, while for ${\bf v}={\bf v}^1_{\rm soft}$ this is true only for $q>q_{un}$. For large values of $q$ the barrier along the perturbed paths lies below $\epsilon_{\rm th}$, but this is true only within the range $q_{ms}<q<q_M$, when the arrival point is a rank-1 saddle. We find that this remains true for arbitrary values of  $\epsilon_a < \epsilon_{\rm th}$. For the largest $q \lesssim q_M$, the curve associated to ${\bf v}^1_{\rm soft}$ is flat, indicating that the energy profile becomes monotonically increasing in the interval $\gamma \in [0,1]$ with a maximum at $\gamma=1$, equal to $\epsilon_1$. 
In view of this comparison, one also understands why the geodesic path is no longer optimal when ${\bf v} \to {\bf v}_{\rm soft}^1$ and $u^1_{\rm typ} \neq 0$: in this case, indeed, ${\bf v}_{\rm soft}^1$ has an $\mathcal{O}(1)$ projection on the vector ${\bf v}_{\rm Hess}$, see Appendix \ref{app:gradient_max_barr}, and thus allowing the path to deviate in the direction of ${\bf v}_{\rm soft}^1$ leads to a lower energy barrier. \\

\emph{Discussion.} The motivation for our work is characterizing paths connecting minima in rough energy landscapes. By focusing on a prototypical mean-field glass model, we have found that the energy profile along the geodesic path connecting two local minima always reaches values of energies that are above the threshold one;
moreover, interpolating paths that are allowed to deviate from the geodesic along 
the direction associated to the softest curvature of the landscape at ${\bm \sigma}^0$ are on average associated to higher energy barriers. Therefore, for the $3$-spin model the direction of softest local curvature at the departing minimum ${\bm \sigma}^0$ is not a predictor of lower energy barriers, and it is likely to be uncorrelated to good transition paths. 
This is different from what observed in numerical simulations of finite dimensional jammed and mildly super-cooled particles \cite{xu2010anharmonic,widmer2008irreversible, khomenko2021relationship}. A possible explanation is that those cases were focusing on portions of the energy landscapes in which indeed the Hessian plays a crucial role \cite{wyart2005geometric,broderix2000energy,angelani2000saddles,coslovich2019localization}: in the former case the energy configurations were drawn close to the jamming transition, whereas in the latter close to the Mode-Coupling cross-over. We focus instead in a regime which would correspond to a deep super-cooled state. Our results therefore suggest that in this case the Hessian plays a different role - a result that would be interesting to test by studying energy paths and barriers in small systems \cite{heuer2008exploring,baity2021revisiting}.\\
On the other hand, the Hessian matrix at ${\bm \sigma}^0$ can still be used to construct directions in configuration space along which to deform the geodesic path, in order to lower the energy barrier: this special direction is obtained acting with the Hessian matrix on the vector that is tangent to the geodesic path at its initial point, and projecting the resulting vector in the subspace orthogonal to the tangent. \\
When the stationary points are closer and more strongly correlated, the direction of softest curvature at ${\bm \sigma}^1$ corresponds to an isolated mode of the Hessian;  this case of nearby, correlated minima is likely to be the one more closely related to the results found in finite dimensional systems \cite{coslovich2019localization}: in fact, we find that in this setting deformation of the geodesic path along the direction of the isolated mode \emph{does} lead to lower energy barriers. The eigenvector associated to the isolated mode is localized in the direction connecting the two local minima (even though one can not talk about quasi-localized modes in this framework of fully-connected systems). The barriers of the perturbed paths lie in some cases below the threshold energy: we find that this happens only when  ${\bm \sigma}^1$ is a rank-1 saddle. 
For smaller $q$, paths lying below the threshold connecting the two minima might exist, but the curvature at ${\bm \sigma}^a$ is not enough to identify them.  \\
Our results shed some light on the challenging problem of activated glassy dynamics \cite{rizzo2021path}, and in particular the identification of the barriers crossed by typical dynamical paths.
The energy barriers found along the interpolating paths are in fact a possible proxy for such barriers, and they give some indications on how the dynamical transition rates between minima depend on the parameters $q, \epsilon_a$.
At very low temperature, one expects that the first energy barriers crossed by the system in its dynamics starting from a metastable state ${\bm \sigma}^0$ correspond to the rank-1 saddles in the landscape that are closer to ${\bm \sigma}^0$~\cite{auffinger2013random, auffinger2020number, ros2019complexity, subag2017complexity,ros2020distribution}. These saddles connect the reference minimum to other local minima that are quite close to the original one\footnote{For any of the rank-1 saddles that lie at $q>q_{ms}$, one can determine the properties of the local minima that are connected to the reference one ${\bm \sigma^0}$ by this saddle \cite{ros2021dynamical}. These local minima typically are at overlap $q \in [q_{\rm un}, q_{\rm ms}]$ with ${\bm \sigma}^0$. We see that the interpolating path between these minima and the reference one have a local maximum that is at  values of energies that do not correspond to that of the saddles, but are higher. Therefore, to reach the rank-1 saddle connecting the pair of minima one has to modify the interpolating path by including other directions in configuration space.}, are correlated to it (i.e., which belong to the region $q \in [q_{un}, q_{ms}]$) and lie at much higher energy~\cite{ros2021dynamical}. The analysis of the interpolating paths complements these findings, as it provides some information on the energy barriers between distant minima at smaller energy difference, that are likely to be connected to the first one  
 by a non-trivial sequence of saddles, i.e. of activated jumps. It is clear that further information can be obtained by identifying  the distribution of rank-1 saddles that connect two \emph{given} local minima ${\bm \sigma}^a$ by means of a direct counting (i.e., by performing a doubly constrained complexity calculation); we defer this analysis to future work.  It could also be interesting to extend this study to mixed models obtained summing terms of the form \eqref{eq:landscape} with different values of $p$ (such as $p=3$ and $p=4$), in particular looking at paths interpolating between the different families of marginally stable minima that are exhibited by those models \cite{folena2020rethinking,kent2023arrangement}: this is also left to future work.


\begin{acknowledgments}
We thank Stephane d'Ascoli for useful discussions at the earlier stage of this work.
VR acknowledges funding by the ``Investissements d’Avenir” LabEx PALM (ANR-10-LABX-0039-PALM). GB acknowledges funding from the Simons Foundation (\#454935 Giulio Biroli)

\end{acknowledgments}

\bibliography{refs.bib}

\onecolumngrid
\appendix

\newpage
\vspace{1 cm}
\section{Tangent planes, Riemannian gradients and Hessians}
\label{app:geometrical_context}

In this Appendix, we introduce the main formalism and notation used in the rest of the paper. Consider the energy field  $h({\bm\sigma})$, defined on the $N$-dimensional hypersphere of unit radius, ${\bm\sigma}\in\mathcal{S}_N(1)$. We introduce a symmetric random tensor with components $a_{i_1 \cdots i_p}$ (which are mean zero and unit variance random variables) so that
\begin{equation}
h({\bm \sigma})= \sqrt{p!} \sum_{i_1 <  \cdots <  i_p} a_{i_1 \cdots i_p} {\bm \sigma}_{i_1} \cdots {\bm \sigma}_{i_p}=\frac{1}{\sqrt{p!}} \sum_{i_1, \cdots, i_p} a_{i_1 \cdots i_p} {\bm \sigma}_{i_1} \cdots {\bm \sigma}_{i_p}= \sqrt{\frac{2}{N}} \mathcal{E}(\sqrt{N} {\bm \sigma}).
\end{equation}
We define with $\nabla h({\bm \sigma})$ the $N$-dimensional gradient of the field $h({\bm\sigma})$ extended to the space $\mathbb{R}^N$. Such vector has components:
\begin{equation}
  (\nabla h({\bm \sigma}))_i =  {\bf x}_i \cdot  \nabla h({\bm \sigma})= \frac{\partial h({\bm \sigma})}{\partial \sigma_i},
\end{equation}
where ${\bf x}_i$ with $i=1, \cdots, N$ denotes some orthonormal basis of $\mathbb{R}^N$ with components $[{\bf x}_i]_j= \delta_{ij}$. Similarly, we define with $\nabla^2 h({\bm \sigma})$ the $N \times N$-dimensional Hessian matrix with components:
\begin{equation}
  (\nabla^2 h({\bm \sigma}))_{ij} =  {\bf x}_i \cdot  \nabla^2 h({\bm \sigma}) \cdot {\bf x}_j= \frac{\partial^2 h({\bm \sigma})}{\partial \sigma_i \partial \sigma_j}.
\end{equation}
Notice that, due to the homogeneity of the field $ h({\bm \sigma})$, one can show that: 
\begin{align}
    \label{eq:app:first_grad_id}
    \nabla h({\bm \sigma})\cdot{\bm\sigma}=p \, h({\bm \sigma}),
\end{align}
and similarly
\begin{align}
    \label{eq:app:hes_to_grad}
    \nabla^2h({\bm \sigma})\cdot{\bm \sigma}=(p-1)\nabla h({\bm \sigma}).
\end{align}

The Riemannian gradient $\bf g({\bm \sigma})$ and Hessian $\mathcal{H}({\bm \sigma})$  of $h({\bm\sigma})$ take into account the spherical constraint imposed on ${\bm \sigma}$. We denote with $\tau[{\bm\sigma}]$ the $(N-1)$-dimensional tangent plane for each point ${\bm\sigma}$ of the hypersphere, which is the plane in $\mathbb{R}^N$ spanned by vectors that are orthogonal to the vector ${\bm \sigma}$ (see Fig. \ref{fig:SketchPath}). Let   $\mathbf{e}_{i<N}({\bm\sigma})$ be an orthonormal basis of such tangent plane, 
\begin{align*}
\tau[{\bm\sigma}]=\text{Span}\{\mathbf{e}_i({\bm\sigma})\}_{i=1}^{N-1}.
\end{align*}
This local basis of $\tau[{\bm\sigma}]$ can be extended to a basis of $\mathbb{R}^N$ by adding $\mathbf{e}_N({\bm\sigma}):={\bm\sigma}$. The $(N-1)$-dimensional Riemannian gradient 
$\mathbf{g}({\bm\sigma})$ is defined, component-wise,  as the projection of the unconstrained gradient $\nabla h({\bm \sigma})$ on the basis elements of the tangent plane  $\tau[{\bm\sigma}]$:
\begin{equation}
{\bf g}_\alpha({\bm\sigma})=\nabla h({\bm \sigma})\cdot {\bf e}_{\alpha}({\bm\sigma}) \quad \text{for} \quad\alpha<N.
\end{equation}
With a slight abuse of notation, in the following we sometimes use the same notation $\mathbf{g}$ for the  $N$-dimensional extended vector with an additional last component ${\bf g}_N({\bm\sigma})=0$.
In this way, the unconstrained and Riemannian gradients are related by:
\begin{align}
    \label{eq:app:second_grad_id}
    \nabla h({\bm \sigma})= {\bf g}({\bm\sigma})+p\, h({\bm\sigma}){\bm\sigma}.
\end{align}

In order to obtain the Riemannian Hessian one can make use of a Lagrange multiplier to enforce the constraint ${\bm\sigma}^2=1$: we  define $h_{\lambda}({\bm\sigma}):=h({\bm\sigma})-\frac{\lambda}{2}({\bm\sigma}^2-1)$. Then $\nabla h_\lambda({\bm\sigma})=\nabla h({\bm \sigma})-\lambda{\bm\sigma}\overset{!}{=}0\Rightarrow \lambda={\bm\sigma}\cdot\nabla h({\bm \sigma})$. From this it follows that
\begin{align}
\label{eq:app:hessian}
\mathcal{H}_{\alpha \beta}({\bm\sigma}):= \mathbf{e}_{\alpha}({\bm\sigma})\cdot \nabla^2 h_\lambda({\bm\sigma}) \cdot \mathbf{e}_{\beta}({\bm\sigma})=\mathbf{e}_{\alpha}({\bm\sigma})\cdot\nabla^2 h({\bm \sigma})\cdot \mathbf{e}_{\beta}({\bm\sigma})-(\nabla h(\bm\sigma)\cdot\bm\sigma)\delta_{\alpha \beta},\quad\quad \alpha,\beta \leq N-1.
\end{align}
The identity \eqref{eq:app:first_grad_id} implies that the Riemannian Hessian is obtained from the unconstrained one by shifting by a diagonal matrix proportional to $p \, h({\bm \sigma})$, and projecting on the local tangent plane. Therefore, working with the unconstrained or  Riemannian Hessian is essentially the same, provided that one remembers about the shift.\\

The tangent plane $\tau[{\bm\sigma}]$, and therefore the basis vectors $\mathbf{e}_{\alpha}({\bm\sigma})$, depend on the particular point in configuration space that one is looking at. For two different configurations, ${\bm \sigma}^0$ and  ${\bm \sigma}^1$ at overlap $q={\bm \sigma}^0\cdot {\bm \sigma}^1$, it is convenient to choose the bases on the tangent planes $\tau[{\bm \sigma}^a]$, with $a=0,1$, as follows: first, one can always choose the basis ${\bf x}_i$
in the $N$-dimensional space $\mathbb{R}^N$ in such a way that the first $N-2$ vectors ${\bf x}_1, \cdots, {\bf x}_{N-2}$ are orthogonal to both ${\bm \sigma}^0$ and ${\bm \sigma}^1$. These vectors belong to both tangent planes (because they are orthogonal to both the configuration vectors), and therefore one can choose ${\bf e}_i({\bm \sigma}^a)= {\bf x}_i$ for all $i=1, \cdots, N-2$. Concretely, in the ${\bf x}_i$ basis one can set ${\bm \sigma}^0=(0, 0, \cdots, 0,1)$ and ${\bm \sigma}^1=(0,0, \cdots, -\sqrt{1-q^2},q)$. Then, in the tangent plane $\tau[{\bm \sigma}^0]$ there remains a single  basis vector to be chosen, which will have a non-zero projection with ${\bm \sigma}^1$; this vector has already been defined in Eq.\eqref{eq:def_basis},
\begin{align*}
 {\bf e}_{N-1}({\bm \sigma}^0)=\frac{q {\bm \sigma}^0-{\bm \sigma}^1}{\sqrt{1-q^2}}.
\end{align*}
It has unit norm and it is orthogonal to all the others, since ${\bf  x}_i \perp {\bm \sigma}^a$ for any $i$ and $a=0,1$. The vector ${\bm \sigma}^0$ then completes the local basis. With the choice above, $ {\bf e}_{N-1}({\bm \sigma}^0)=(0,0, \cdots, 1, 0)$. Similarly, $\tau[{\bm \sigma}^1]$ is spanned by ${\bf x}_1, \cdots, {\bf x}_{N-2}$ plus the vector ${\bf e}_{N-1}({\bm \sigma}^1)$ defined similarly as
\begin{align*}
 {\bf e}_{N-1}({\bm \sigma}^1)=\frac{q {\bm \sigma}^1-{\bm \sigma}^0}{\sqrt{1-q^2}}.
\end{align*}

Again, with the choice above $ {\bf e}_{N-1}({\bm \sigma}^1)=(0,0, \cdots, -q, -\sqrt{1-q^2})$.
In summary, we can choose ${\bf e}_\alpha ({\bm \sigma}^0)={\bf e}_\alpha ({\bm \sigma}^1)$  with  $\alpha=1, \cdots, N-2$ as a basis of the subspace  $\text{span}\{{\bf x}_\alpha\}$, while the remaining basis vector of the tangent planes ${\bf e}_{N-1} ({\bm \sigma}^a)$ changes depending on which tangent plane we consider (in Sec.~\ref{app:QuenchedDistribution}, we specify the choice of some of the vectors ${\bf e}_\alpha ({\bm \sigma}^0)={\bf e}_\alpha ({\bm \sigma}^1)$  as well).  We denote the sets of these basis vectors as $\mathcal{B}^a= \left\{ {\bf e }_\alpha({\bm \sigma}^a)\right\}_{\alpha=1}^N$. All along this work, we assume that this choice of basis is made. Moreover, we will often use the notation ${\bf e}_\alpha ({\bm \sigma}^a)={\bf e}_\alpha ^a$, as well as $h({\bm \sigma}^a)=h^a$,  ${\bf g}_\alpha({\bm\sigma}^a)={\bf g}_\alpha^a$ and $\mathcal{H}({\bm\sigma}^a)= \mathcal{H}^a$ for simplicity.

\section{The complexity calculation: a recap}
\label{app:recap_paper_saddles}
For completeness, we provide a summary of the main results of Ref.~\cite{ros2019complexity}, upon which part of the present work is built. In Ref.~\cite{ros2019complexity} the authors compute the quenched complexity of stationary points ${\bm\sigma}^1$ of $h({\bm \sigma})$ at energy density $\epsilon_1$ (meaning that ${\bf g}({\bm\sigma}^1)=0$ and $h({\bm\sigma}^1)=\sqrt{2 N} \epsilon_1$), that are conditioned to be at fixed overlap ${\bm \sigma}^0 \cdot {\bm \sigma}^1= q$ with a given \emph{reference} minimum ${\bm\sigma}_0$ at energy density $\epsilon_0$. We refer to ${\bm \sigma}^1$ as the \emph{secondary} configuration. The complexity reads
\begin{align*}
\Sigma(\epsilon_1,q|\epsilon_0)=\lim_{N\to\infty}\frac{1}{N}\mathbb{E}\left[ \log\mathcal{N}_{{\bm\sigma}_0}(\epsilon_1,q|\epsilon_0)
    \right],
\end{align*}
see Eq.~\eqref{eq:quenchedcomp}. The expectation value indicates both an average  
over the disorder, as well as a flat average over  all local minima of energy $\epsilon_0$; the random variable $\mathcal{N}_{{\bm\sigma}^0}(\epsilon_1,q|\epsilon_0)$ is the number of such stationary points ${\bm\sigma}^1$, and it is defined below in Eq.~\eqref{eq:app:counting_functions}.  To compute the expected value of the logarithm by means of the replica trick, one needs to replicate the secondary configuration ${\bm\sigma}^1$:
\begin{align}
\Sigma(\epsilon_1,q|\epsilon_0)=\lim_{N\to\infty}\lim_{n\to 0} \frac{M_n(\epsilon_1,q|\epsilon_0)-1}{Nn}
\end{align}
with
\begin{align}\label{eq:moments}
M_n(\epsilon_1,q|\epsilon_0):=\mathbb{E}\left[
\frac{1}{\mathcal{N}(\epsilon_0)}\int_{\mathcal{S}_N(1)}d{\bm\sigma}^0\,\omega({\bm\sigma}^0)\int_{\mathcal{S}_N(1)}\prod_{a=1}^n d{\bm\sigma}^a\omega_{\epsilon_1,q}({\bm\sigma^a}|{\bm\sigma}^0)
\right]
\end{align}
and 
\begin{align}
    \begin{split}
        &\omega_{\epsilon_0}({\bm \sigma}^0)=|\text{det}\mathcal{H}({\bm \sigma}^0)|\,\,\delta \tonde{h({\bm \sigma}^0)-\sqrt{2N}\epsilon_0} \,\, \delta \tonde{{\bf g}({\bm \sigma}^0) }\\
        &\omega_{\epsilon_1,q}({\bm \sigma}^a|{\bm \sigma}^0)=|\text{det}\mathcal{H}({\bm \sigma}^a)|\,\,\delta(h({\bm \sigma}^a)-\sqrt{2N}\epsilon_1)\,\, \delta \tonde{{\bf g}({\bm \sigma}^a) }\,\,\delta({\bm \sigma}^0\cdot{\bm \sigma}^a-q) 
    \end{split}
\end{align}
and where
\begin{equation}
\label{eq:app:counting_functions}
    \begin{split}
        \mathcal{N}(\epsilon_0)&= \int_{\mathcal{S}_N(1)} d{\bm \sigma}^0\, \omega_{\epsilon_0}({\bm \sigma}^0 ), \\
        \mathcal{N}_{{\bm \sigma}^0}(\epsilon_1, q|\epsilon_0)&=\int_{\mathcal{S}_N(1)} d{\bm \sigma}^1 \,\omega_{\epsilon_1,q}({\bm \sigma}^1| {\bm \sigma}^0)
    \end{split}
\end{equation}
represent, respectively, the number of stationary points at energy density $\epsilon_0$ and the number of stationary points at energy density $\epsilon_1$ and overlap $q$ with ${\bm \sigma}^0$. In order to carry out such computation, in principle one has to replicate the reference configuration ${\bm\sigma}^0$ as well by raising the denominator to the numerator, thus writing:
\begin{align}
M_n(\epsilon_1,q|\epsilon_0):=\lim_{k\to 0}\mathbb{E}\left[
\int_{\mathcal{S}_N(1)}\prod_{\beta=1}^k d{\bm\sigma}^{0,\beta}\,\omega({\bm\sigma}^{0,\beta})\int_{\mathcal{S}_N(1)}\prod_{a=1}^n d{\bm\sigma}^a\omega_{\epsilon_1,q}({\bm\sigma^a}|{\bm\sigma}^0),
\right]
\end{align}
where ${\bm\sigma}^{0,\beta=1}= {\bm\sigma}^{0}$.
Due to the isotropy of the correlation function of the random field $h({\bm \sigma})$, it turns out that this expectation value depends on the configurations ${\bm \sigma}^a, {\bm \sigma}^{0,\beta}$ only through the overlaps $q_{\alpha \beta}^0={\bm \sigma}^{0,\alpha} \cdot {\bm \sigma}^{0,\beta}$, $ q_{a \beta}={\bm \sigma}^{a} \cdot {\bm \sigma}^{0,\beta}$  and $q^1_{a b}={\bm \sigma}^{a} \cdot {\bm \sigma}^{b}$, where by construction $q_{\alpha \alpha}^0= q^1_{aa}=1$ and $ q_{a 1}={\bm \sigma}^{a} \cdot {\bm \sigma}^{0,\beta=1}=q$. This implies that the integral over the configurations can be replaced by an integral over these order parameters, which can be computed with the saddle point method. One sees that the saddle point equations for the parameters $q_{\alpha \beta}^0$ enforce $q_{\alpha \beta}^0=0$ for all $\alpha \neq \beta=1, \cdots, k$; this reflects the fact that the overlap between metastable states in the pure spherical $p$-spin model is vanishing \cite{crisanti1992sphericalp}, i.e. the states are typically orthogonal to each others. This also implies that $ q_{a \beta}=0$ for all $a=1, \cdots n$ and $\beta \neq 1$. As a consequence, at the saddle point solution the secondary replicas ${\bm \sigma}^1$ are coupled only to the original reference configuration  ${\bm \sigma}^0$ and not to its replicas. It is simple to check that this implies that to leading order in $N$, the expectation value \eqref{eq:moments} is identical to its annealed version, which is obtained averaging separately the numerator and the denominator. 
 This means that we can write
\begin{align}
\begin{split}
M_n(\epsilon_1,q|\epsilon_0):&=\frac{1}{\mathbb{E}[\mathcal{N}(\epsilon_0)]}\mathbb{E}\left[
\int_{\mathcal{S}_N(1)}d{\bm\sigma}^{0}\,\omega({\bm\sigma}^{0})\int_{\mathcal{S}_N(1)}\prod_{a=1}^n d{\bm\sigma}^a\omega_{\epsilon_1,q}({\bm\sigma^a}|{\bm\sigma}^0)
\right]\\
&=\mathbb{E}\left[
\int_{\mathcal{S}_N(1)}\prod_{a=1}^n d{\bm\sigma}^a\omega_{\epsilon_1,q}({\bm\sigma^a}|{\bm\sigma}^0)\bigg| \;  \begin{subarray}{l}
 h({\bm \sigma}^0) = \sqrt{2 N}\epsilon_0\\ 
 {\bf g}({\bm \sigma}^0)={\bf 0} \end{subarray}\right]\\
 &=\int_{\mathcal{S}_N(1)}\prod_{a=1}^n d{\bm\sigma}^a\,\mathbb{E}\left[\prod_{a=1}^n\omega_{\epsilon_1,q}({\bm\sigma}^a|{\bm\sigma}^0)\bigg| \;  \begin{subarray}{l}
 h({\bm \sigma}^0) = \sqrt{2 N}\epsilon_0\\ 
 {\bf g}({\bm \sigma}^0)={\bf 0} \end{subarray}\right]\\
&=\int_{\mathcal{S}_N(1)}\prod_{a=1}^n d{\bm\sigma}^a\,\mathbb{E}\left[\prod_{a=1}^n \,   |\text{det} \mathcal{H}({\bm \sigma}^a)|  \; \Big| \;   \grafe{
 \begin{subarray}{l}
 h^a = \sqrt{2 N}\epsilon_1, h^0=\sqrt{2 N} \epsilon_0\\ 
 {\bf g}^a={\bf 0}, \, {\bf g}^0={\bf 0} \; \forall  a=1, ...,n \end{subarray}}\right]  p_{\vec{{\bm \sigma}}|{\bm \sigma}^0}({\bf 0}, \epsilon_1)
\end{split}
\end{align}
where $\vec{\bm \sigma}=({\bm \sigma}^1, \cdots, {\bm \sigma}^n)$ and 
\begin{align}
\label{eq:app:jpdf}
p_{\vec{{\bm \sigma}}|{\bm \sigma}^0}({\bf 0}, \epsilon_1)=\mathbb{E}\quadre{
\prod_{a=1}^n\delta(h^a-\sqrt{2N}\epsilon_1) \delta( {\bf g}^a) \; \Big| \;  \begin{subarray}{l}
 h^0=\sqrt{2 N} \epsilon_0\\ 
 {\bf g}^0={\bf 0} \end{subarray}}.
\end{align}
The expectation value is now a function of the overlaps $q^1_{ab}$, and its leading order term can be determined again with a saddle point calculation. Searching for a saddle point solution where all replicas have the same overlap, $q^1_{ab}= q_1$ for all $a,b=1, \cdots n$ and $a \neq b$, one finds the solution $q_1=q^2$ \cite{ros2019complexity}, which is the smallest possible overlap between replicas that are all subject to the constraint of having overlap $q$ with the reference configuration. It can be showed explicitly that this implies that the complexity $\Sigma(\epsilon_1,q|\epsilon_0)$ computed within the quenched formalism is the same as that obtained within the annealed framework, i.e., setting $n=1$ in the formulas above. More precisely, one obtained the identity:
\begin{align*}\label{eq:QuenchedEqualAnnealed}
\Sigma(\epsilon_1,q|\epsilon_0)=\lim_{N\to\infty}\frac{1}{N}\mathbb{E}\left[ \log\mathcal{N}_{{\bm\sigma}_0}(\epsilon_1,q|\epsilon_0) \right] = \lim_{N\to\infty}\frac{1}{N}\log \mathbb{E}\left[ \mathcal{N}_{{\bm\sigma}_0}(\epsilon_1,q|\epsilon_0).
    \right]
\end{align*}

The final formula giving the complexity reads \cite{ros2019complexity}:
\begin{align}
    \Sigma(\epsilon_1,q|\epsilon_0)=\frac{1}{2}\log\left(\frac{p}{2}(\tilde{z}-\epsilon_1)^2\right)+\frac{p(\epsilon_1^2+\epsilon_1\tilde{z})}{2(p-1)}+\frac{Q}{2}, \quad \quad \tilde{z}=\sqrt{\epsilon_1^2-\epsilon_{\rm th}^2}
\end{align}
where $\epsilon_{\rm th}=- \sqrt{2(p-1)/p}$ and 
\begin{align}
    Q=\log\left(\frac{1-q^2}{1-q^{2p-2}}\right)-2(\epsilon_0^2U_0(q)+\epsilon_0\epsilon_1U(q)+\epsilon_1^2U_1(q))
\end{align}
with
\begin{equation}
 \begin{split}
  U_0(q)&=\frac{q^{2 p} \left[-q^{2 p}+p q^2 \left(1-q^2\right)+q^4\right]}{q^{4 p}-q^{2 p} \left[(p-1)^2 (1+q^4)-2 p(p-2) q^2\right] +q^4},\\
  U(q)&=\frac{2 q^{3 p} \left[1-p \left(1-q^2\right)\right]-2 q^{p+4}}{q^{4 p}-\left((p-1)^2 (1+q^4)-2 (p-2) p q^2\right) q^{2 p}+q^4},\\
  U_1(q)&=\frac{q^4-q^{2 p}-p q^{2 p}\left[(p-1) q^4+(3-2 p) q^2+p-2\right]}{q^{4 p}-q^{2 p} \left[(p-1)^2 (1+q^4)-2 p(p-2) q^2\right] +q^4}.
 \end{split}
\end{equation}

\begin{figure}[h!]
\centering
\includegraphics[width=0.5\textwidth]{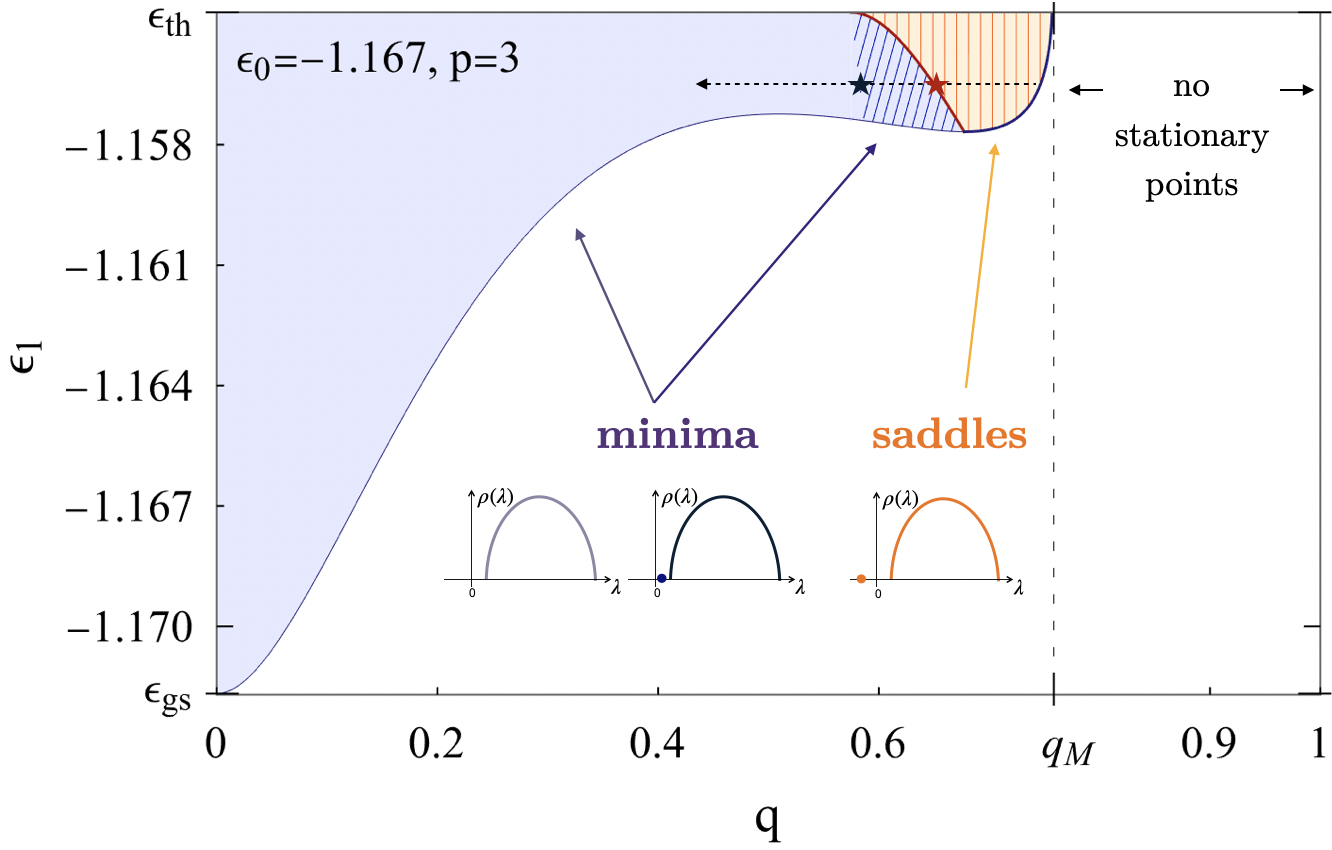}
\caption{The plot shows in color the region in the plane $(q, \epsilon_1)$ where the complexity is positive, for $\epsilon_0=-1.167$. The blue leftmost area corresponds to minima, the dashed blue zone to minima with an isolated eigenvalue in the Hessian spectrum, the rightmost (dashed) yellow zone to index-1 saddles. The stars mark the transitions in the properties of the Hessians of the stationary points, for fixed $\epsilon_1$. In the white area the complexity is vanishing. }
\label{app:fig:sigma_1}
\end{figure}

The results of this calculation are summarized in  Fig.~\ref{app:fig:sigma_1} for a representative value of $\epsilon_0=-1.167$ ~\cite{ros2021dynamical}.
For fixed $\epsilon_0< \epsilon_{\rm th}$, one finds that $\Sigma(\epsilon_1,q|\epsilon_0)$ is positive only for $q$ smaller than a threshold value $q_M$, which depends on $\epsilon_0$. For fixed $q< q_M$, the complexity is positive for $\epsilon_1$ belonging to a finite range, that is shown as a colored area (blue and yellow) in Fig.~\ref{app:fig:sigma_1}. By looking at the statistics of the Hessian of the stationary points at given $q, \epsilon_1$, one can discriminate whether they are stable local minima (such that all the eigenvalues of the Hessian are typically positive) or saddles with index $k$, i.e. with an Hessian with $k$ negative eigenvalues. As we recall below, the statistic of the Hessian matrices is the same as that of GOE matrices perturbed with additive and multiplicative low-rank perturbations, and shifted by the quantity $p \epsilon_1$. Fixing the energy $\epsilon_1$ and decreasing the overlap $q$ (see the dashed arrow in Fig.~\ref{app:fig:sigma_1}), one finds that the typical stationary points are first saddles of index $k=1$ (yellow area), then minima with an isolated but positive mode (blue dashed area), and finally minima with an Hessian spectrum described simply by a shifted semicircular law, without any isolated mode (blue area). The transitions between these different populations of stationary points are marked with stars in the figure. For $\epsilon> \epsilon_{\rm th}$, all stationary points are saddles of extensive index $k \sim \mathcal{O}(N)$.

In the following, we select pairs of stationary points with parameters $\epsilon_a, q$ which belong to the colored region, i.e. which are such that the corresponding complexity $\Sigma(\epsilon_1, q|\epsilon_0)\geq  0$.

\section{The energy profile along a path, and the case $p=3$}
\label{app:energy_density_profile}

Consider a path on the hypersphere, parametrized as in Eq.~\eqref{eq:path}. Let $h[\gamma, f] \equiv h({\bm \sigma}[\gamma, f]) $ denote the  energy profile along the path. It holds:
\begin{equation}
    h[\gamma, f]=\frac{1}{\sqrt{p!}} \sum_{i_1, \cdots, i_p} a_{i_1 \cdots i_p} \sum_{k_1=0}^p \sum_{k_2=0}^{k_1} \binom{p}{k_1} \binom{k_1}{k_2} f^{p-k_1} \beta^{k_1-k_2} \gamma^{k_2} {\bf v}_{i_1} \cdots {\bf v}_{i_{p-k_1}} {\bm \sigma}^0_{i_{p-k_1+1}} \cdots {\bm \sigma}^0_{i_{p-k_2}}{\bm \sigma}^1_{i_{p-k_2+1}} \cdots {\bm \sigma}^0_{i_{p}}.
\end{equation}
In the special case $p=3$, this function depends only on the gradients and the Hessians of the energy field at the initial and final configurations ${\bm \sigma}^a$, $a=0,1$. Indeed, for $p=3$ the above expansion reduces to:
\begin{equation}\label{eq:app:energy_profile0}
\begin{split}
    h[\gamma,f] \stackrel{p=3}{=}\, &\gamma^3 h({\bm \sigma}^1)+ \beta^3 h({\bm \sigma}^0)+ f^3 h({\bf v})+  \gamma^2 \beta \; \nabla h({\bm \sigma}^1)\cdot {\bm \sigma}^0+  \beta^2 \gamma \; \nabla h({\bm \sigma}^0)\cdot {\bm \sigma}^1
 +\gamma^2 f \; \nabla h({\bm \sigma}^1)\cdot {\bf v} +  \beta^2 f \; \nabla h({\bm \sigma}^0)\cdot \bf{v}
\\
&+ \gamma \beta f  \; {\bf v}\cdot  \nabla^2 h({\bm \sigma}^0) \cdot {\bm \sigma}^1+   \frac{f^2}{2} \tonde{  \gamma \, {\bf v} \cdot \nabla^2 h({\bm \sigma}^1)\cdot {\bf v} + \beta\, {\bf v} \cdot \nabla^2 h({\bm \sigma}^0)\cdot {\bf v}}.
\end{split}
\end{equation}
We are interested in the case in which  the beginning and end points of the path are stationary points satisfying ${\bf g}({\bm \sigma}^a)=0$, and ${\bm \sigma}^0 \cdot {\bm \sigma}^1=q$. Moreover, we assume that the vector ${\bf v}$ is chosen in such a way that ${\bf v}\perp {\bm \sigma}^0,{\bm \sigma}^1$. The identity \eqref{eq:app:second_grad_id} then implies that 
\begin{align}
\label{eq:app:energy_profile}
\begin{split}
       h[\gamma,f]=
&\tonde{\gamma^3 +3\gamma^2 \beta \,q} h({\bm \sigma}^1)+ \tonde{\beta^3+3 \beta^2 \gamma \,q} h({\bm \sigma}^0)+ f^3 h({\bf v})+ \gamma \beta f  \; {\bf v}\cdot  \nabla^2 h({\bm \sigma}^0) \cdot {\bm \sigma}^1\\
& +   \frac{f^2}{2} \tonde{  \gamma \, {\bf v} \cdot \nabla^2 h({\bm \sigma}^1)\cdot {\bf v} + \beta\, {\bf v} \cdot \nabla^2 h({\bm \sigma}^0)\cdot {\bf v}}.
\end{split}
\end{align}
From this equation, Eq. \eqref{eq:PrePAth} follows by using the block structure of the conditional Hessian matrices described below, as well as the fact that $h({\bf v})=0$ typically. We remark that this expression is at fixed random couplings. In the rest of the paper, we restrict to the case $p=3$ and average the energy profile over all configurations ${\bm \sigma}^0$ and ${\bm \sigma}^1$ satisfying
\begin{align}
\label{eq:app:constraints}
    \begin{split}
    &{\bf g}({\bm \sigma}^0)=\textbf{{\bf 0}}\quad \&\quad {\bf g}({\bm \sigma}^1)=\textbf{{\bf 0}}\\
&h({\bm \sigma}^0)=\sqrt{2N}\epsilon_0\quad\&\quad h({\bm \sigma}^1)=\sqrt{2N}\epsilon_1\\
    &{\bm \sigma}^0\cdot{\bm \sigma}^1=q,
    \end{split}
\end{align}
as well as over the distribution of the random couplings $a_{i_1\ldots i_p}$.

\section{Averages: quenched vs annealed energy profiles}
\label{app:setup}
 
\noindent It follows from Eq.~\eqref{eq:app:energy_profile} that the energy profile along a path is completely determined by the local properties at the two configurations ${\bm \sigma}^a$ for $a=0,1$, if ${\bf v}$ is also chosen to be a local property that depends only in the statistics of the landscape at  ${\bm \sigma}^0$ or  ${\bm \sigma}^1$. In the following, we assume that  ${\bf v}$ is correlated with the softest mode of the Hessian matrix at either  ${\bm \sigma}^0$ or ${\bm \sigma}^1$.  We now consider the average of this function, over the family of stationary points ${\bm \sigma}^a$ of energy density $\epsilon_a$ and overlap $q$, and over all possible realizations of the random landscape. We consider two different averaging prescriptions, which we refer to as \emph{quenched} and \emph{annealed} in analogy with \cite{franz1998effective}.  \\

{\bf The \emph{quenched} energy profile. }
The quenched averaging protocol corresponds to the following procedure: (i) for fixed realization of the random landscape, select ${\bm \sigma}^0$ among the exponentially many stationary points of energy density $\epsilon_0$; (ii) select ${\bm \sigma}^1$ among the stationary points of energy density $\epsilon_1$ that are at overlap $q$ with the previously selected  ${\bm \sigma}^0$; (iii) evaluate the energy profile along the path, under the condition that ${\bf v}$ has a projection on the minimal Hessian eigenvector at the start or at the end; (iv) average the profile over the population of stationary points ${\bm \sigma}^1$ with parameters $\epsilon_1, q$, (v) average over all ${\bm \sigma}^0$, (vi) average over the realizations of the landscape. We recall that in order to satisfy the spherical constraint, all integrals over configurations are taken over $\mathcal{S}_N(1)$. Formally, this prescription corresponds to:
\begin{equation}\label{eq:AvQ}
h^{(Q)}[\gamma, f]:=\mathbb{E}\quadre{\frac{1}{\mathcal{N}(\epsilon_0)} \int_{\mathcal{S}_N(1)} d{\bm \sigma}^0\, \omega_{\epsilon_0}({\bm \sigma}^0 ) \frac{1}{\mathcal{N}_{{\bm \sigma}^0}(\epsilon_1, q|\epsilon_0)} \int_{\mathcal{S}_N(1)} d{\bm \sigma}^1 \, \omega_{\epsilon_1,q}({\bm \sigma}^1| {\bm \sigma}^0) \, h[\gamma,f]}
\end{equation}
Where all the quantities above have already been defined in Sec.~\ref{app:recap_paper_saddles}. In \eqref{eq:AvQ}, the random landscape appears both in the numerator and in the denominator; in order to perform the average, one has to resort to the replica trick to treat both denominators, through the formula $x^{-1}=\lim_{n \to 0} x^{n-1}$. Since there are two denominators, two sets of replicas should be exploited. However, due to the fact that the typical overlap between stationary points ${\bm \sigma}^0$ in the pure spherical $p$-spin is zero \cite{crisanti1992sphericalp}, the flat average over ${\bm \sigma}^0$ can be reproduced by conditioning the average over the landscape to realizations for which the configuration  ${\bm \sigma}^0$ is stationary, with a given energy density $\epsilon_0$ (see Sec.~\ref{app:recap_paper_saddles}). 
 This implies:
\begin{equation}\label{eq:app:Planted}
h^{(Q)}[\gamma, f]=\mathbb{E}\quadre{ \frac{1}{\mathcal{N}_{{\bm \sigma}^0}(\epsilon_1, q|\epsilon_0)} \int_{\mathcal{S}_N(1)} d{\bm \sigma}^1 \omega_{\epsilon_1,q}({\bm \sigma}^1| {\bm \sigma}^0) \, h[\gamma,f] \; \Big| \;  \begin{subarray}{l}
 h({\bm \sigma}^0) = \sqrt{2 N}\epsilon_0\\ 
 {\bf g}({\bm \sigma}^0)={\bf 0} \end{subarray}}.
\end{equation}
By means of the replica trick, this average can be re-written as:
\begin{equation}
h^{(Q)}[\gamma, f]=\lim_{n \to 0} \,  \int_{\mathcal{S}_N(1)} \prod_{k=1}^n d{\bm \sigma}^k   \; \mathbb{E}\quadre{ \prod_{k=1}^n\,  \omega_{\epsilon_1,q}({\bm \sigma}^k| {\bm \sigma}^0) \, h[\gamma,f] \; \Big| \;  \begin{subarray}{l}
 h({\bm \sigma}^0) = \sqrt{2 N}\epsilon_0\\ 
 {\bf g}({\bm \sigma}^0)={\bf 0} \end{subarray}}.
\end{equation}
The delta functions inside $\omega_{\epsilon_1,q}({\bm \sigma}^1|{\bm \sigma}^0)$ can be replaced by a conditioning in the average. To simplify the notation, for all the $n+1$ configurations ${\bm \sigma}^k$  with $k=0,1,\cdots,n$ we define the gradient vectors ${\bf g}^k \equiv {\bf g}({\bm \sigma}^k)$, the Hessian matrices $\mathcal{H}^k \equiv \mathcal{H}({\bm \sigma}^k)$, and the rescaled energy functional $h^k\equiv h({\bm \sigma}^k)$. This leads to:
\begin{equation}\label{eq:app:final_average}
h^{(Q)}[\gamma, f]=\lim_{n \to 0} \,  \int_{\mathcal{S}_N(1)} \prod_{k=1}^n d{\bm \sigma}^k   \,\delta \tonde{{\bm \sigma}^k \cdot {\bm \sigma}^0- q}\, \mathbb{E}\quadre{\prod_{k=1}^n \,   |\text{det} \mathcal{H}({\bm \sigma}^k)|  \, h[\gamma,f] \; \Big| \;   \grafe{
 \begin{subarray}{l}
 h^k = \sqrt{2 N}\epsilon_1, h^0=\sqrt{2 N} \epsilon_0\\ 
 {\bf g}^k={\bf 0}, \, {\bf g}^0={\bf 0} \; \forall  k=1, ...,n \end{subarray}} } p_{\vec{{\bm \sigma}}|{\bm \sigma}^0}({\bf 0}, \epsilon_1),
\end{equation}
where $\vec{\bm \sigma}=({\bm \sigma}^1, \cdots, {\bm \sigma}^n)$ and $p_{\vec{{\bm \sigma}}|{\bm \sigma}^0}({\bf 0}, \epsilon_1)$, the joint density function of the gradients $\vec{\bf g}^k$ and fields $\vec{h}^k$, induced by the distribution of the couplings  and conditioned to ${\bf g}^0={\bf 0}$ and $h^0= \sqrt{2 N} \epsilon_0$, have already been defined in Eq.~\eqref{eq:app:jpdf}

We define:
\begin{equation}
      \mathbb{E}^{\rm cond}_n\quadre{\; \cdot \; } \equiv \mathbb{E}\quadre{\;\cdot \; \Big| \;   \grafe{
 \begin{subarray}{l}
 h^k = \sqrt{2 N}\epsilon_1, h^0=\sqrt{2 N} \epsilon_0\\ 
 {\bf g}^k={\bf 0}, \, {\bf g}^0={\bf 0} \; \forall  k=1, ...,n \end{subarray}} }.
\end{equation}
Plugging \eqref{eq:app:energy_profile} into \eqref{eq:app:final_average} and implementing the conditioning, we find:
\begin{equation}
\label{eq:app:QuenchedProf}
\begin{split}
&h^{(Q)}[\gamma, f]=\tonde{\gamma^3 +3\gamma^2 \beta \,q} \sqrt{2N} \epsilon_1+ \tonde{\beta^3+3 \beta^2 \gamma \,q}\sqrt{2N} \epsilon_0+\\
& f^3(\gamma) \lim_{n \to 0} \,  \int_{\mathcal{S}_N(1)} \prod_{k=1}^n d{\bm \sigma}^k   \,\delta \tonde{{\bm \sigma}^k \cdot {\bm \sigma}^0- q}\, \mathbb{E}^{\rm cond}_n\quadre{\prod_{k=1}^n \,   |\text{det} \mathcal{H}({\bm \sigma}^k)|  \,h({\bf v}) \, } p_{\vec{{\bm \sigma}}|{\bm \sigma}^0}({\bf 0}, \epsilon_1) +\\
&\gamma \, \beta[\gamma; f]\, f(\gamma)\, \lim_{n \to 0} \,  \int_{\mathcal{S}_N(1)} \prod_{k=1}^n d{\bm \sigma}^k   \,\delta \tonde{{\bm \sigma}^k \cdot {\bm \sigma}^0- q}\, \mathbb{E}^{\rm cond}_n\quadre{\prod_{k=1}^n \,   |\text{det} \mathcal{H}({\bm \sigma}^k)|  \,{\bf v}\cdot  \nabla^2 h({\bm \sigma}^0) \cdot {\bm \sigma}^1 \, } p_{\vec{{\bm \sigma}}|{\bm \sigma}^0}({\bf 0}, \epsilon_1) +\\
&  \frac{f^2(\gamma)}{2} \lim_{n \to 0} \,  \int_{\mathcal{S}_N(1)} \prod_{k=1}^n d{\bm \sigma}^k   \,\delta \tonde{{\bm \sigma}^k \cdot {\bm \sigma}^0- q}\, \mathbb{E}^{\rm cond}_n\quadre{\prod_{k=1}^n \,   |\text{det} \mathcal{H}({\bm \sigma}^k)|  \, \tonde{  \gamma \, {\bf v} \cdot \nabla^2 h({\bm \sigma}^1)\cdot {\bf v} + \beta\, {\bf v} \cdot \nabla^2 h({\bm \sigma}^0)\cdot {\bf v}}\, } p_{\vec{{\bm \sigma}}|{\bm \sigma}^0}({\bf 0}, \epsilon_1),
\end{split}
\end{equation}
where the Riemannian and unconstrained Hessian matrices are related by \eqref{eq:app:hessian}.
The above equation contains the conditional average of the product of $n$ determinants, multiplied by those terms in the energy profile $h$ that depend on the matrices $\mathcal{H}({\bm\sigma}^1)$ and $\mathcal{H}({\bm \sigma}^0)$. All the matrices $\mathcal{H}({\bm  \sigma}^\alpha)$ with $\alpha=0,1,\cdots,n$ are coupled, and each one has to be conditioned to the gradients and energies of the $n+1$ configurations. The correct way to  proceed is to determine the expectation value for fixed $n$, and subsequently take $n \to 0$. In the language of the Franz-Parisi potential \cite{franz1998effective}, this is a \emph{quenched} calculation, since the two configurations ${\bm \sigma}^0$ and ${\bm \sigma}^1$ are not on the same footing: ${\bm \sigma}^0$ can be considered as a reference configuration that is selected first, and ${\bm \sigma}^1$ as a secondary one.  \\

{\bf The \emph{annealed} energy profile. } In the annealed case, the two configurations  ${\bm \sigma}^0$ and  ${\bm \sigma}^1$ are treated on the same footing; the averaging protocol corresponds to the following procedure: (i) for fixed realization of the random landscape, select ${\bm \sigma}^0$ and ${\bm \sigma}^1$ among the exponentially many stationary points of energy density $\epsilon_0$ and $\epsilon_1$, at mutual overlap $q$; (ii) proceed as in the quenched case. Formally, this is equivalent to taking Eq.~\eqref{eq:app:Planted} and factorizing the average of the numerator and denominator, getting: 
\begin{equation}
h^{(A)}[\gamma, f]= \frac{1}{\mathbb{E}\quadre{\mathcal{N}_{{\bm \sigma}^0}(\epsilon_1, q|\epsilon_0)}}\int_{\mathcal{S}_N(1)}  d{\bm \sigma}^1   \,\delta \tonde{{\bm \sigma}^1 \cdot {\bm \sigma}^0- q}\, \mathbb{E}\quadre{   |\text{det} \mathcal{H}({\bm \sigma}^1)|  \, h[\gamma,f] \; \Big| \;   \grafe{
 \begin{subarray}{l}
 h^1 = \sqrt{2 N}\epsilon_1, h^0=\sqrt{2 N} \epsilon_0\\ 
 {\bf g}^\alpha={\bf 0} \; \forall  \alpha=0, 1 \end{subarray}} } p_{{\bm \sigma}^1|{\bm \sigma}^0}({\bf 0}, \epsilon_1),
\end{equation}
where now
\begin{align}
p_{{\bm \sigma}^1|{\bm \sigma}^0}({\bf 0}, \epsilon_1)=\mathbb{E}\quadre{\delta(h^1-\sqrt{2N}\epsilon_1) \delta( {\bf g}^1)  \; \Big| \;  \begin{subarray}{l}
 h^0=\sqrt{2 N} \epsilon_0\\ 
 {\bf g}^0={\bf 0} \end{subarray}}.
\end{align}
One sees that the numerator is the same as \eqref{eq:app:final_average} with $n=1$. Plugging \eqref{eq:app:energy_profile} into this expression and implementing the conditioning, we find in this case:

\begin{equation}\label{eq:app:Annealed2}
\begin{split}
&h^{(A)}[\gamma, f]= \tonde{\gamma^3 +3\gamma^2 \beta \,q} \sqrt{2N}\epsilon_1+ \tonde{\beta^3+3 \beta^2 \gamma \,q} \sqrt{2N} \epsilon_0+\\
&\frac{f^3(\gamma)}{\mathbb{E}\quadre{\mathcal{N}_{{\bm \sigma}^0}(\epsilon_1, q|\epsilon_0)}}  \,  \int_{\mathcal{S}_N(1)}  d{\bm \sigma}^1  \,\delta \tonde{{\bm \sigma}^1 \cdot {\bm \sigma}^0- q}\, \mathbb{E}^{\rm cond}_1\quadre{ \,   |\text{det} \mathcal{H}({\bm \sigma}^1)|  \,h({\bf v}) \, } p_{{\bm \sigma}^1|{\bm \sigma}^0}({\bf 0}, \epsilon_1) +\\
&\frac{\gamma \, \beta[\gamma; f]\, f(\gamma)}{\mathbb{E}\quadre{\mathcal{N}_{{\bm \sigma}^0}(\epsilon_1, q|\epsilon_0)}}  \, 
  \int_{\mathcal{S}_N(1)}  d{\bm \sigma}^1   \,\delta \tonde{{\bm \sigma}^1 \cdot {\bm \sigma}^0- q}\, \mathbb{E}^{\rm cond}_1\quadre{ \,   |\text{det} \mathcal{H}({\bm \sigma}^1)|  \,{\bf v}\cdot  \nabla^2 h({\bm \sigma}^0) \cdot {\bm \sigma}^1 \, } p_{{\bm \sigma}^1|{\bm \sigma}^0}({\bf 0}, \epsilon_1) +\\
&\frac{f^2(\gamma)}{2 \mathbb{E}\quadre{\mathcal{N}_{{\bm \sigma}^0}(\epsilon_1, q|\epsilon_0)}}  \, 
   \int_{\mathcal{S}_N(1)}  d{\bm \sigma}^1   \,\delta \tonde{{\bm \sigma}^1 \cdot {\bm \sigma}^0- q}\, \mathbb{E}^{\rm cond}_1\quadre{   |\text{det} \mathcal{H}({\bm \sigma}^1)|  \, \tonde{  \gamma \, {\bf v} \cdot \nabla^2 h({\bm \sigma}^1)\cdot {\bf v} + \beta\, {\bf v} \cdot \nabla^2 h({\bm \sigma}^0)\cdot {\bf v}}\, } p_{\vec{{\bm \sigma}}|{\bm \sigma}^0}({\bf 0},\epsilon_1).
\end{split}
\end{equation}
In this case, one has to deal with a pair of  matrices, conditioned to only two gradients and energies. In the language of the Franz-Parisi potential, this is an \emph{annealed} calculation.

\section{Statistics of the Hessian matrices}
\label{app:statistics_HESSIANS}
In this Appendix, we discuss the statistical distribution of the unconstrained Hessian matrices  $\frac{1}{\sqrt{N-1}}\nabla^2 h({\bm \sigma}^a)$ with $a=0,1$, subject to the conditioning on the energies and gradients of the various replicas. As we have seen in Appendix \ref{app:geometrical_context}, the unconstrained Hessian and the Riemannian Hessian are easily related by a projection and shift.

\subsection{Statistics of the Hessians: the annealed setup}\label{app:AnnealedDistribution}

\subsubsection{Matrix distribution after conditioning}
In this case, one has to determine the joint distributions of the two Hessian matrices $\nabla^2 h({\bm \sigma}^a)$, each one conditioned to  ${\bf g}({\bm \sigma}^a)={\bf 0}$ and $h({\bm \sigma}^a)= \sqrt{2 N} \epsilon_a$ for $a=0,1$. This conditional, joint distribution has been determined in \cite{ros2019complexity,subag2017complexity}: each matrix $\nabla^2 h({\bm \sigma}^a)$, expressed in its own tangent plane basis $\mathcal{B}^a$ introduced in Appendix \ref{app:geometrical_context}, has the following block structure:   
\begin{equation}
\label{eq:app:annealed_matrix_0}
\frac{1}{\sqrt{N-1}}\nabla^2 h({\bm \sigma}^a) = \begin{pmatrix}
 & & &m^a_{1 N-1}&0\\
 \\
& {\bf B}^a&&\vdots &0\\
\\
 & & &m^a_{N-2 \,N-1}&0\\
m^a_{1 N-1}& \cdots &m^a_{N-2 \,N-1}&m^a_{N-1 \,N-1}+\mu_a&0\\
0&0&0&0&  \sqrt{\frac{2 N}{N-1}} \, p(p-1)\epsilon_a
\end{pmatrix}\in\mathbb{R}^{N\times N}.
\end{equation}
The entries in the $(N-2) \times (N-2)$ blocks ${\bf B}^a$ are independent of the entries $m^a_{i N-1}$, and correlated with each others as follows:
\begin{equation}\label{eq:b_correlations}
\begin{split}
   & \mathbb{E}[B_{ij}^a \, B_{kl}^b]= \tonde{\delta_{a b} \frac{\sigma^2}{N-1}+ (1-\delta_{ab})\frac{\sigma^2_H}{N-1}}(\delta_{ik} \delta_{jl}+ \delta_{il} \delta_{jk}), \quad \quad \\
    & \mathbb{E}[m_{iM}^a \, m_{jM}^b]= \tonde{\delta_{a b} \frac{\Delta^2}{N-1}+ (1-\delta_{ab})\frac{\Delta^2_h}{N-1}}\delta_{ik} \quad \quad i,j <N-1, \\
    & \mathbb{E}[m_{N-1N-1}^a m_{N-1N-1}^b]= \frac{v_{ab}}{N-1}, \quad \quad a, b \in \{ 0,1 \}.
    \end{split}
\end{equation}

Therefore, the blocks ${\bf B}^a$ are $(N-2) \times (N-2)$ matrices with GOE (Gaussian Orthogonal Ensemble) statistics, with rescaled variance $\sigma^2 (N-2)/(N-1)$. The two blocks are coupled component-wise. The parameters $\sigma, \sigma_H, \Delta, \Delta_h$ and $\mu_a$ are functions of $p$ and of the parameters $\epsilon_a, q$, and read \cite{ros2019complexity}: 

\begin{equation}\label{eq:app:FormPar}
\begin{split}
&\sigma^2=p(p-1)\\
&\sigma_H^2=p(p-1)q^{p-2}\\
&\sigma_W^2=\sigma^2-\sigma_H^2=p(p-1)(1-q^{p-2})\\
&\Delta^2=p(p-1) \quadre{1-\frac{(p-1)(1-q^2) q^{2p-4}}{1-q^{2p-2}}}\\
&\Delta_h^2=p(p-1) q^{p-3} (-1)\quadre{1-\frac{(p-1) (1-q^2)}{1-q^{2p-2}}}\\
&\mu_1= \mu(\epsilon_0, \epsilon_1), \quad \mu_0= \mu(\epsilon_1, \epsilon_0) 
\end{split}
\end{equation}
where: 
\begin{equation}\label{eq:app:Mu}
 \begin{split}
&\mu(x, y) \equiv \sqrt{2} p(p-1) \left(1-q^2\right)\frac{ [q^4- (p-1)q^{2 p}+ (p-2)q^{2 p+2}] x-[q^{3 p}+(p-2) q^{p+2}-(p-1) q^{p+4}] y }{q^{6-p}+q^{3 p+2}- q^{p+2}[(p-1)^2 (1+q^4)-2 p (p-2) q^2]}.\\
 \end{split}
\end{equation}
The conditional distribution of the Riemannian Hessians therefore reads as follows:
\begin{equation}
\label{eq:app:annealed_matrix_1a}
\frac{\mathcal{H}({\bm \sigma}^a)}{\sqrt{N-1}} = \begin{pmatrix}
 & & &m^a_{1 N-1}\\
 \\
& {\bf B}^a&&\vdots \\
\\
 & & &m^a_{N-2 \,N-1}\\
m^a_{1 N-1}& \cdots &m^a_{N-2 \,N-1}&m^a_{N-1 \,N-1}
\end{pmatrix}+ \mu_a \,{\bf e}_{N-1}^a [{\bf e}_{N-1}^a]^T -\sqrt{\frac{2 N}{N-1}} \, p\epsilon_a \quad \in\mathbb{R}^{N-1\times N-1}.
\end{equation}
For later convenience, we also define the matrix:
\begin{equation}
\label{eq:app:TildeMatDef} 
\frac{\tilde{\mathcal{H}}_A({\bm \sigma}^a)}{\sqrt{N-1}} = \begin{pmatrix}
 & & &m^a_{1 N-1}\\
 \\
& {\bf B}^a&&\vdots \\
\\
 & & &m^a_{N-2 \,N-1}\\
m^a_{1 N-1}& \cdots &m^a_{N-2 \,N-1}&m^a_{N-1 \,N-1}+\mu^a
\end{pmatrix} \quad \in\mathbb{R}^{N-1\times N-1}.
\end{equation}
where $A$ stands for annealed. These matrices can be though of as matrices with a GOE statistics, perturbed with both an additive and multiplicative finite rank perturbation along the direction corresponding to the basis vector ${\bf e}_{N-1}({\bm \sigma}^a)$, and shifted by a diagonal matrix. Notice that in the annealed formalism the random matrix problem is symmetric, in the sense that the two conditioned Hessians have the same structure at both the configurations ${\bm \sigma}^0$ and ${\bm\sigma}^1$. \\

Let us now consider specifically the case $p=3$. One sees from \eqref{eq:app:FormPar} that for $p=3$ it holds
$\Delta^2=\Delta_h^2= 6 (1-q^2)(1+q^2)^{-1}$,
which implies that $m^0_{iN-1}=m^1_{iN-1}\equiv m_{iN-1}$. Moreover, in this case $\mu_1= 6 \sqrt{2} q \tonde{\epsilon_0-q \epsilon_1}(1-q^2)^{-1}$ and $\mu_0= 6 \sqrt{2} q \tonde{\epsilon_1-q \epsilon_0}(1-q^2)^{-1}$.

\subsubsection{Spectral statistics and isolated eigenvalue}
The matrices \eqref{eq:app:annealed_matrix_1a} have a block structure, with a $(N-2) \times (N-2)$ block ${\bf B}^a$ with GOE statistics, and a special line and column. Let us neglect the term proportional to the identity, which corresponds to a constant shift of the whole spectrum, and work with $\tilde{\mathcal{H}}_A({\bm\sigma^a})$.
The spectral properties of these kind of matrices are discussed in detail in Ref. \cite{paccoros}. To leading order in the size of the matrix, the eigenvalue density $\rho_N^a(\lambda)$ of both matrices is not affected  by the presence of the special line and column, and it just coincides with the the eigenvalue density of the GOE block, i.e., it is given by the semicircular law
\begin{equation}
\rho_N^a(\lambda) =\rho_\sigma(\lambda)+ \mathcal{O}\tonde{\frac{1}{N}}=\frac{1}{2 \pi \sigma^2} \sqrt{4 \sigma^2 - \lambda^2}+ \mathcal{O}\tonde{\frac{1}{N}}, \quad \quad \sigma^2=p(p-1).
\end{equation}

 The presence of the special row and column can give rise to subleading contributions to the eigenvalue density: these contributions correspond to eigenvalues that do not belong to the support of the semicircular law (and are said to be ``isolated"), and whose typical value depends on the parameters $\Delta, \mu_a$ governing the statistics of the entries of the special row and column. As argued in Ref.~\cite{paccoros}, the fact that $\Delta \leq \sigma$ (as can be easily verified to be the case here), implies that only one isolated eigenvalue can exist for these matrices. Such eigenvalue exists whenever 
\begin{equation}
\label{eq:app:ConditionEx}
|\mu_a|>  \sigma \tonde{1+ \frac{\sigma^2-\Delta^2}{\sigma^2}},
\end{equation}
and it is obtained as the real solution of the equation
\begin{equation}
\label{eq:app:egval_eqn}
    \lambda - \mu_a - \Delta^2 \mathfrak{g}_\sigma(\lambda)=0,
\end{equation}
where for $z$ real such that $ |z|> 2 \sigma$ one has
\begin{equation}
\mathfrak{g}_\sigma(z)=\frac{1}{2 \sigma^2} \tonde{z-  \text{sign}(z)\sqrt{z^2- 4 \sigma^2}}. 
\end{equation}
The equation \eqref{eq:app:egval_eqn} is obtained imposing that the resolvent of the matrix \eqref{eq:app:TildeMatDef}, projected into the direction ${\bf e}^a_{N-1}$ of the rank-1 perturbation, has a pole. Explicitly, the eigenvalue is given by:
\begin{equation}
\label{eq:app:IsoExplicit}
\lambda_{\rm iso}^a=\frac{{2 \mu_a \sigma^2- \Delta^2 \mu_a- \text{sign}(\mu_a) \Delta^2 \sqrt{\mu_a^2-4 (\sigma^2- \Delta^2)}}}{2 (\sigma^2-\Delta^2)}.
\end{equation}
Notice that for $\mu_a <0$, this eigenvalue is negative and it coincides with the smallest one of the matrix, i.e., $\lambda_{\rm \min}^a= \lambda_{\rm iso}^a$. 
Whenever the isolated eigenvalue exists, its eigenvector ${\bf e}^a_{\rm iso}$ has a projection on the vector ${\bf e}_{N-1}({\bm\sigma}^a)$, corresponding to the special line and column of the matrix, which remains of $\mathcal{O}(1)$ when $N$ is large. The typical value of this projection has been computed in Refs.~\cite{ros2020distribution, paccoros} and reads:
\begin{equation}
\label{eq:app:ProjVectors}
\begin{split}
  &({\bf e}^a_{\rm iso} \cdot  {\bf e}_{N-1}^a)^2= \mathfrak{q}_{\sigma,\Delta}(\lambda_{\rm iso}^a,\mu_a)
   \end{split}
\end{equation}
where we introduced the function:
\begin{equation}\label{eq:Qfunc}
\begin{split}
   & \mathfrak{q}_{\sigma,\Delta}(\lambda,\mu):=\text{sign}(\mu)\,\frac{\text{sign}(\lambda)\Delta^2\sqrt{\lambda^2-4\sigma^2}-\lambda(2\sigma^2-\Delta^2)+2\mu\sigma^2}{2\Delta^2\sqrt{\mu^2-4(\sigma^2-\Delta^2)}}.
    \end{split}
\end{equation}
It is shown in Ref.~\cite{paccoros} that Eq.~\eqref{eq:app:ProjVectors} is indeed positive on the Right Hand Side, as it should be. Moreover, whenever the squared overlap is non-zero, the eigenvector is aligned with the direction of the finite-rank perturbation, meaning that ${\bf e}^a_{\rm iso} \cdot  {\bf e}_{N-1}^a>0$. In the main text, we consider the case in which $\mu_a<0$ and thus $\lambda_{\rm \min}^a= \lambda_{\rm iso}^a$, and define $u^a= ({\bf e}^a_{\rm iso} \cdot  {\bf e}_{N-1}^a)^2$. The expression for these quantities in the case $a=1$ and $p=3$ are given in Eq.~\eqref{eq:Typicals}. 

\subsubsection{Change of basis}
 We recall that the matrices \eqref{eq:app:annealed_matrix_0} are expressed in the bases $\mathcal{B}^a$. It is sometimes convenient for the calculations to express both matrices in the same basis. This is achieved by performing a simple rotation, which involves only the two elements of the basis sets. We use the fact that 
\begin{equation}\label{eq:Idi}
\begin{split}
&{\bm \sigma}^0=q{\bm \sigma}^1-\sqrt{1-q^2}{\bf e}_{N-1}({\bm \sigma}^1)\quad\quad {\bf e}_{N-1}({\bm \sigma}^0)=-\sqrt{1-q^2}{\bm \sigma}^1-q{\bf e}_{N-1}({\bm \sigma}^1)\\
&{\bm \sigma}^1=q{\bm \sigma}^0-\sqrt{1-q^2}{\bf e}_{N-1}({\bm \sigma}^0)\quad\quad {\bf e}_{N-1}({\bm \sigma}^1)=-\sqrt{1-q^2}{\bm \sigma}^0-q{\bf e}_{N-1}({\bm \sigma}^0).
\end{split}
\end{equation}

The matrix of change of basis $\mathcal{B}^0 \to \mathcal{B}^1$ reads:
\begin{equation}
\label{eq:app:ChangeBas}
{\bf R}_{0 \to 1}= \begin{pmatrix}
1 & 0&  \ldots& 0&0&0\\
&\ddots&  &&0&0\\
&&\ddots  &&0&0\\
0 & \ldots& 0&1 &0&0\\
0&\ldots&\ldots&0&-q&-\sqrt{1-q^2}\\
0& \ldots&\ldots& 0&-\sqrt{1-q^2}&  q
\end{pmatrix}\in\mathbb{R}^{N\times N},
\end{equation}
and it is equivalent to ${\bf R}_{1 \to 0}$. This implies that in the basis $\mathcal{B}^1$ we have:
\begin{align}
\label{eq:app:hess_new_basis_ann}
\begin{split}
&{\bf R}_{0 \to 1}^{-1} \frac{\nabla^2 h({\bm \sigma}^0)}{\sqrt {N-1}}   {\bf R}_{0 \to 1}
\\&= \begin{pmatrix}
 & &  & &- q m^0_{1N-1}&-\sqrt{1-q^2}m^0_{1N-1}\\
 \\
&& {\bf B}^0 &&-q m^0_{iN-1}& -\sqrt{1-q^2}m^0_{iN-1} \\
\\
 & &  & &- q m^0_{N-2 N-1}& -\sqrt{1-q^2} m^0_{N-2 N-1}\\
-q m^0_{1 N-1}&\cdots&\cdots&-q m^0_{N-2 N-1 }&q^2( m^0_{N-1 N-1}+\mu_0)+l^0_{N-1 N-1}&l^0_{N-1 N}\\
- \sqrt{1-q^2} m^0_{1N-1} & \cdots& \cdots& -\sqrt{1-q^2} m^0_{N-2 N-1} &l^0_{ N-1 N}& l^0_{NN}
\end{pmatrix},
\end{split}
\end{align}
where 
\begin{equation}
\label{eq:app:SpecialComp}
\begin{split}
&l_{N-1 N-1}^0=  (1-q^2) \sqrt{\frac{2 N}{N-1}} p (p-1) \epsilon_0,\\
&l_{NN}^0=(1-q^2) (m^0_{N-1N-1}+\mu_0) + q^2  \sqrt{\frac{2 N}{N-1}} p (p-1) \epsilon_0,\\
&l_{N-1 N}^0= q \sqrt{1-q^2} \tonde{m^0_{N-1N-1}+\mu_0 -\sqrt{\frac{2 N}{N-1}}p(p-1) \epsilon_0 }.
\end{split}
\end{equation}

Defining the $(N-1) \times (N-1)$ matrix
\begin{equation}
    \Pi_q^a=\mathbb{I}-(1-q){\bf e}_{N-1}({\bm \sigma}^a){\bf e}_{N-1}^T({\bm \sigma}^a)
\end{equation}
and using \eqref{eq:app:TildeMatDef}, one can write the projection of this matrix onto the basis of $\tau[{\bm \sigma}^1]$ as 
\begin{align}
\label{eq:app:BasisChanged_annealed}
    {\bf R}_{0 \to 1}^{-1} \, \frac{\nabla^2 h({\bm \sigma}^0)}{\sqrt{N-1}}\, {\bf R}_{0 \to 1}\bigg|_\perp=
    \Pi_q^1
    \;\cdot \frac{\tilde{\mathcal{H}}_A({\bm \sigma}^0)}{\sqrt{N-1}}\cdot\;
    \Pi_q^1+l_{N-1 N-1}^0{\bf e}_
    {N-1}({\bm \sigma}^1){\bf e}_{N-1}^T({\bm \sigma}^1)
\end{align}
Notice that the matrix $\nabla^2h({\bm \sigma}^1)$ expressed in the basis $\mathcal{B}^0$ has exactly the same form, with the superscript $0$ replaced by $1$.

\subsection{Statistics of the Hessians: the quenched setup}\label{app:QuenchedDistribution}

\subsubsection{Matrix distribution after conditioning}
In the quenched setting, replicas are introduced and one has to consider the joint distribution of $n+1$ matrices $\nabla^2h({\bm \sigma}^\alpha)$ with $\alpha=0, 1, \cdots, n$, conditioned to ${\bf g}({\bm \sigma}^\alpha)={\bf 0}$ and $h({\bm \sigma}^0)= \sqrt{2 N} \epsilon_0$, $h({\bm \sigma}^k)= \sqrt{2 N} \epsilon_1$ with $k=1, \cdots, n$. The joint distribution of the matrices $\nabla^2h({\bm \sigma}^k)$ with $k=1, \cdots, n$ subject to the above conditioning is derived in Ref.~\cite{ros2019complexity}. We recall it in the following, discussing in addition the statistics of $\nabla^2h({\bm \sigma}^0)$ and its correlations with $\nabla^2h({\bm \sigma}^1)$, which are the two matrices appearing in the energy profile. In this section, $M:=N-n$.\\

The conditional Hessians have a block structure when expressed in an appropriate basis. We make a particular choice of the vectors ${\bf e}_i({\bm \sigma}^a)$ for $i=M, \cdots, N-2$. As in Appendix \ref{app:recap_paper_saddles}, we define the overlap between the replicas as
\begin{equation}
\label{eq:app:Overlap}
    q_1= {\bm \sigma}^k \cdot {\bm \sigma}^l \quad \quad k,l =1, \cdots, n, \quad \quad k \neq l.
\end{equation}
For $a=1$, we choose the basis vectors in $\mathcal{B}^1$ as: 
\begin{equation}
    \begin{split}
        &{\bf e}_{M}({\bm \sigma}^1)=\frac{{\bm \sigma}^2-{\bm \sigma}^3}{\sqrt{2 (1-q_1)}}\\
          &{\bf e}_{M+1}({\bm \sigma}^1)=\frac{{\bm \sigma}^2+{\bm \sigma}^3-2{\bm \sigma}^4}{\sqrt{3 \cdot 2 (1-q_1)}}\\
            &{\bf e}_{M+2}({\bm \sigma}^1)=\frac{{\bm \sigma}^2+{\bm \sigma}^3+{\bm \sigma}^4-3 {\bm \sigma}^5}{\sqrt{4 \cdot 3 (1-q_1)}}\\
            &\cdots\\
              &{\bf e}_{N-3}({\bm \sigma}^1)=\frac{{\bm \sigma}^2+ \cdots +{\bm \sigma}^{n-1}-(n-2){\bm \sigma}^{n} }{\sqrt{(n-1)  (n-2) (1-q_1)}}\\
               &{\bf e}_{N-2}({\bm \sigma}^1)=\frac{{\bm \sigma}^2+ \cdots +{\bm \sigma}^n-(n-1) q_1{\bm \sigma}^1+ (n-1) q \frac{1-q_1}{1-q^2}(q {\bm \sigma}^1-{\bm \sigma}^0 ) }{\sqrt{(n-1) \quadre{1+(n-2) q_1-(n-1)q_1^2 - (n-1)q^2 \frac{1-q_1}{1-q^2} (1-q_1 ) }}}
    \end{split}
\end{equation}
and, as specified in Appendix \ref{app:geometrical_context}, 
\begin{align*}
&{\bf e}_{N-1}({\bm \sigma}^1)=\frac{q {\bm \sigma}^1-{\bm \sigma}^0}{\sqrt{1-q^2}}.
\end{align*}
It is simple to check that all these vectors are all orthogonal to each others under the assumption \eqref{eq:app:Overlap}, and are orthogonal to ${\bm \sigma}^1$: therefore, they all belong to the tangent plane $\tau[{\bm \sigma}^1]$. When using the notation $\mathcal{B}^1$ in the following, we assume this choice of basis vectors. 
Notice that the typical value of the overlap \eqref{eq:app:Overlap} between stationary points (as it follows from the saddle-point calculation of the complexity recalled in Appendix \ref{app:recap_paper_saddles}) is $q_1= q^2$, thus at the saddle point:
\begin{equation}
    \begin{split}
              &{\bf e}_{N-2}({\bm \sigma}^1)
           \stackrel{q_1=q^2}{=} \frac{{\bm \sigma}^2+ \cdots +{\bm \sigma}^n-(n-1) q{\bm \sigma}^0 }{\sqrt{(n-1) \quadre{1-q^2 }}}.
    \end{split}
\end{equation}

The choice of the basis vectors for ${\bm\sigma}^0$ is analogous:
\begin{equation}
 {\bf e}_{i}({\bm \sigma}^1)={\bf e}_{i}({\bm \sigma}^0) \quad \quad i=M, \cdots, N-3
\end{equation}
with 
\begin{equation}
\begin{split}
    {\bf e}_{N-2}({\bm \sigma}^0)=\frac{{\bm \sigma}^2+ \cdots +{\bm \sigma}^n-(n-1) q{\bm \sigma}^0+ (n-1)  \frac{q_1-q^2}{1-q^2}(q {\bm \sigma}^0-{\bm \sigma}^1 ) }{\sqrt{(n-1) \quadre{1-q^2 + (n-2) (q_1- q^2)-(n-1)(1-q^2) \tonde{\frac{q_1-q^2}{1-q^2}}^2 - 2 \frac{(q_1-q^2)^2}{1-q^2} }}}
    \end{split}
\end{equation}
and again
\begin{equation}
    \begin{split}
     &{\bf e}_{N-1}({\bm \sigma}^0)=\frac{q {\bm \sigma}^0-{\bm \sigma}^1}{\sqrt{1-q^2}}.
    \end{split}
\end{equation}
Notice that when $q_1=q^2$, the vector ${\bf e}_{N-2}({\bm \sigma}^0)$ simplifies and becomes equivalent to ${\bf e}_{N-2}({\bm \sigma}^1)$, and thus all the basis vectors are the same except for ${\bf e}_{N-1}$. \\

At variance with the annealed case, in the quenched setting the problem is not symmetric in the two configurations ${\bm \sigma}^a$. Consider first the Hessian at ${\bm \sigma^1}$. Its conditional distribution for finite, arbitrary $n$ is given  in Ref.~\cite{ros2019complexity}: in the basis $\mathcal{B}^1$ introduced just above, the unconstrained, conditioned Hessian takes the following form:

\begin{equation}
\frac{\nabla^2 h({\bm \sigma}^1)}{\sqrt{N-1}} = \begin{pmatrix}
 & & &n_{1M}^1 &  \cdots &n_{1 N-2}^1  &m^1_{1N-1} &0\\
 \\
& {\bf B}^1&&\vdots && \vdots&\vdots\\
\\
 &  & &n_{M-1 M}^1 &&  n_{M-1 N-2}^1&m^1_{M-1 N-1} &0\\
n^1_{1\, M}& \cdots&n^1_{M-1\, M}& n^1_{MM}+\nu_1& \cdots&& m^1_{MN-1}&0 \\
\vdots & & &&n^1_{M+1 M+1}+\nu_1 && &0 \\
\vdots  & &&&&\ddots & &0 \\
m^1_{1 \, N-1}& \cdots &m^1_{M-1 \, N-1}& m^1_{M\,N-1}& \cdots&& m^1_{N-1\,N-1}+\mu_1&0 \\
0&0&0&0&0 & 0& 0&\sqrt{\frac{2 N}{N-1}} \, p(p-1)\epsilon_1
\end{pmatrix}
\end{equation}
where ${\bf B}^1$ is a $(M-1) \times (M-1)$ block with the same statistics defined above in the annealed case. 
For $q_1=q^2$ and $i,j<M$ and $ M \leq k,l <N-1 $ it holds: 
\begin{equation}
\mathbb{E}\quadre{n^1_{ik} n^1_{jl}}= \delta_{ij} \delta_{kl} \; p(p-1)\quadre{1- \frac{(p-1)(1-q^2) q^{2p-4}}{1-q^{2p-2}} q^{2p-4}} \equiv \delta_{ij} \delta_{kl} \; \delta^2
\end{equation}
while for $i,j<M$
\begin{equation}
\mathbb{E}\quadre{m^1_{i N-1} m^1_{jN-1}}= \delta_{ij} \; p(p-1)\quadre{1- \frac{(p-1)(1-q^2) q^{2p-4}}{1-q^{2p-2}}}= \delta_{ij} \; \Delta^2,
\end{equation}
where $\Delta^2$ is the same as in \eqref{eq:app:FormPar}. The entries  $n_{ij}^1, m_{iN-1}^1$ with  $ M \leq i,j <N-1 $ form an $n \times n$ block with components that are correlated to each others, but independent from all the other entries of the matrix. We do not report the expression of the corresponding covariances, since it is not needed for the calculation below \footnote{In the following, we are interested in determining the eigenvalue density of the Hessian matrices to order $1/N$, meaning that we aim at determining the bulk of the eigenvalues distribution, as well as the typical value of any isolated eigenvalue that may exist. It turns out that these quantities are insensitive to changes in the variance of $O(N^0)$ components of the matrix, and thus they do not depend on the covariances off the entries of the $n \times n$ block.}. The diagonal terms in this $n \times n$ block contain deterministic contributions that arise from the conditioning, with $\mu_1$ appearing already in \eqref{eq:app:Mu}, and $\nu_1 \neq \mu_1$. The explicit expression of $\nu_1$ is not reported here since it will not be needed in the remainder of the paper, but can be found in Ref.\cite{ros2019complexity} (see Eq. (60) in Appendix F of that paper). As before, one finds that 
the conditional distribution of the Riemannian Hessians in the quenched case reads: 
\begin{equation}
\frac{\mathcal{H}({\bm \sigma}^1)}{\sqrt{N-1}} = \frac{\tilde{\mathcal{H}}_Q({\bm \sigma}^1)}{\sqrt{N-1}} -\sqrt{\frac{2 N}{N-1}} \, p\epsilon_1 \quad \in\mathbb{R}^{N-1\times N-1},
\end{equation}

where now:
\begin{equation}
\label{eq:app:quenched_matrix_1}
\frac{\tilde{\mathcal{H}}_Q({\bm \sigma}^1)}{\sqrt{N-1}} = \begin{pmatrix}
 & & &n_{1M}^1 &  \cdots &n_{1 N-2}^1  &m^1_{1N-1} \\
 \\
& {\bf B}^1&&\vdots &\\
\\
 &  & &n_{M-1 M}^1 &&  n_{M-1 N-2}^1&m^1_{M-1 N-1} \\
n^1_{1\, M}& \cdots&n^1_{M-1\, M}& n^1_{MM}+\nu_1& \cdots&& m^1_{MN-1} \\
\vdots & & &&n^1_{M+1 M+1}+\nu_1 &&  \\
\vdots  & &&&&\ddots &  \\
m^1_{1 \, N-1}& \cdots &m^1_{M-1 \, N-1}& m^1_{M\,N-1}& \cdots&& m^1_{N-1\,N-1}+\mu_1
\end{pmatrix}.
\end{equation}
Here $Q$ stands for quenched. In the quenched case, one has therefore $n$ special lines, $n-1$ out of which have exactly the same statistics controlled by the parameters $\delta^2 $ and $\nu_1$, while the last one has yet different averages $\mu_1$ and variances $\Delta^2$. When $n \to 1$, only the last special line survives and one recovers the annealed Hessian distribution discussed above. \\

We now consider the Hessian at ${\bm \sigma^0}$. Its conditional statistics is not given explicitly in Ref.~\cite{ros2019complexity}, but can be easily derived following the same reasoning of Appendix C of that paper. In the basis $\mathcal{B}^0$ defined above, the unconstrained, conditioned Hessian takes the following form:
\begin{equation}
\label{eq:app:quenched_matrix_0}
\frac{\nabla^2 h({\bm \sigma}^0)}{\sqrt{N-1}} = \begin{pmatrix}
 & & &n_{1M}^0 &  \cdots &n_{1 N-2}^0  &m^0_{1N-1} &0\\
 \\
& {\bf B}^0&&\vdots &\\
\\
 &  & &n_{M-1 M}^0 &&  n_{M-1 N-2}^0&m^0_{M-1 N-1} &0\\
n^0_{1\, M}& \cdots&n^0_{M-1\, M}& n^0_{MM}+\mu_0& \cdots&& m^0_{M\,N-1}&0 \\
\vdots & & &&n^0_{M+1 M+1}+\mu_0 && &0 \\
\vdots  & &&&&\ddots & &0 \\
m^0_{1 \, N-1}& \cdots &m^0_{M-1 \, N-1}& m^0_{M\,N-1}& \cdots&& m^0_{N-1\,N-1}+\mu_0&0 \\
0&0&0&0&0 & 0& 0&\sqrt{\frac{2 N}{N-1}} \, p(p-1)\epsilon_0
\end{pmatrix},
\end{equation}
where ${\bf B}^0$ is again a $(M-1) \times (M-1)$ block with the same statistics defined above in the annealed case, but in this case for $q_1=q^2$ and $i,j<M$ and $ M \leq k,l < N-1 $ it holds: 
\begin{equation}
\mathbb{E}\quadre{n^0_{ik} n^0_{jl}}= \delta_{ij} \delta_{kl} \;  p(p-1)\quadre{1- \frac{(p-1)(1-q^2) q^{2p-4}}{1-q^{2p-2}}}= \delta_{kl} \mathbb{E}\quadre{m^0_{i N-1} m^0_{jN-1}}= \delta_{ij} \delta_{kl} \;\Delta^2.
\end{equation}
Once more, the entries $n_{ij}^0, m_{iN-1}^0$ with  $ M \leq i,j <N-1 $ form an $n \times n$ block with components correlated to each others, but independent from all the other entries of the matrix. Finally, $\mu_0$ is obtained from $\mu_1$ in \eqref{eq:app:Mu} by switching $\epsilon_0 \leftrightarrow \epsilon_1$. As it appears from these formulas, in the case of ${\bm \sigma}^0$ the conditioning modifies the variances of the entries that belong to $n$ special lines and columns of the matrix, in a way that is identical for all rows and columns. Notice that the  covariance of the matrix elements $n^0_{il}$ is different with respect to that of the elements $n^1_{il}$, and it coincides instead with that of the elements $m^a_{i N-1}$. This is consistent with the fact that, for $n \to 1$, one should recover the annealed statistics described in the section above. As before, one finds that 
the conditional distribution of the Riemannian Hessians in the quenched case reads: 
\begin{equation}
\frac{\mathcal{H}({\bm \sigma}^0)}{\sqrt{N-1}} = \frac{\tilde{\mathcal{H}}_Q({\bm \sigma}^0)}{\sqrt{N-1}} -\sqrt{\frac{2 N}{N-1}} \, p\epsilon_1 \quad \in\mathbb{R}^{N-1\times N-1},
\end{equation}
where:
\begin{equation}
\label{eq:app:Quench00}
\frac{\tilde{\mathcal{H}}_Q({\bm \sigma}^0)}{\sqrt{N-1}} = \begin{pmatrix}
 & & &n_{1M}^0 &  \cdots  &m^0_{1N-1} \\
 \\
& {\bf B}^0&&\vdots &\\
\\
 & & &n_{M-1 M}^0 &&m^0_{M-1 N-1} \\
n^0_{1\, M}& \cdots &n^0_{M-1\, M}& n^0_{MM}+ \mu_0& \cdots& m^0_{MN-1} \\
\vdots & & &&&&  \\
m^0_{1 \, N-1}& \cdots &m^0_{M-1 \, N-1}& m^0_{M\,N-1}& \cdots& m^0_{N-1 N-1}+\mu_0
\end{pmatrix}.\\
\end{equation}

Finally, we discuss the correlations between the entries of these two Hessian. This can be obtained with an extension of the calculation described in Appendix C of Ref.~\cite{ros2019complexity}. As recalled in Sec.~\ref{app:recap_paper_saddles}, at the saddle point one finds that the overlap between replicas is $q_1=q^2$. Then, for $i,j<M$ and $ M \leq k,l < N-1 $ it holds: 
\begin{equation}
\mathbb{E}\quadre{n^0_{ik} n^1_{jl}}= \delta_{ij} \delta_{kl} \;  p(p-1)q^{p-2} \quadre{1-(p-1) \frac{q^{2p-4}}{1-q^{2p-2}}(1-q^2)} \equiv \delta_{ij} \delta_{kl} \; \delta_{h}^2,
\end{equation}
while 
\begin{equation}
\mathbb{E}\quadre{m^0_{iN-1} m^1_{jN-1}}= \delta_{ij} \;  p(p-1)q^{p-3}(-1) \quadre{1-(p-1) \frac{1}{1-q^{2p-2}}(1-q^2)} \equiv \delta_{ij} \, \Delta_h^2,
\end{equation}
where $\Delta^2_h$ is the same appearing in \eqref{eq:app:FormPar}. Notice that $\delta^2_h$ differs from $\delta^2$ by a factor of $q^{p-2}$; instead  $\Delta^2_h$ differs from $\Delta^2$ by a factor of $q^{p-3}$, meaning that the two quantities coincide for $p=3$. \\

\subsubsection{Spectral statistics and isolated eigenvalue}
Let us discuss the spectral properties of the conditional Hessian in the case in which  $n$ is an integer parameter. As in the annealed case, we neglect the constant shifts.
In the quenched setting, the distribution of the two matrices is not equivalent: we begin by discussing the shifted Hessian at $ {\bm \sigma}^1$, defined in Eq.~\eqref{eq:app:quenched_matrix_1}. 
This matrix has $(n-1)$ lines with identical statistics given by $\delta^2, \nu_1$, and one last line with statistics controlled by $\Delta^2,\mu_1$. For integer $n$ the isolated eigenvalues, whenever they exist, are solutions of the equation:
\begin{equation}\label{eq:app:egval_eqn_11}
    [\lambda - \nu_1 - \delta^2 \mathfrak{g}_\sigma(\lambda)]^{n-1} \,   [\lambda - \mu_1-\Delta^2 \mathfrak{g}_\sigma(\lambda)]=0,
\end{equation}
which is again obtained imposing that the resolvent of \eqref{eq:app:quenched_matrix_1} projected onto the $n$-dimensional subspace corresponding to the finite-rank perturbations has some poles. One sees that two different eigenvalues may exist, one of which with degeneracy $(n-1)$. In the replica calculation, however, one is required to take the limit $n \to 0$ at the end of the calculation. It appears that in this limit, the first factor of \eqref{eq:app:egval_eqn_11} goes into the denominator, and therefore the corresponding solution becomes a zero, and not a pole, of the resolvent.  
In the limit $n \to 0$, the equation \eqref{eq:app:egval_eqn_11} can exhibit only one meaningful solution, whenever $\lambda - \mu_1-\Delta^2 \mathfrak{g}_\sigma(\lambda)=0$. The corresponding solution is given by \eqref{eq:app:IsoExplicit}, as in the annealed case, and it exists whenever \eqref{eq:app:ConditionEx} is satisfied. Notice that combining this argument with the discussion in the annealed case, one can also conclude that \eqref{eq:app:quenched_matrix_1} subtracted of the last element $\mu_1$ has no isolated eigenvalue in the limit $n \to 0$. 

We now consider the shifted Hessian at 
$ {\bm \sigma}^0$, defined in Eq.~\eqref{eq:app:Quench00}. In this case, there are $n$ special lines with identical statistics. For integer $n$ the isolated eigenvalue, whenever it exists, is the solution of the following equation:
\begin{equation}\label{eq:app:egval_eqn_0}
    [\lambda - \mu_0 - \Delta^2 \mathfrak{g}_\sigma(\lambda)]^{n} =0,
\end{equation}
and has degeneracy $n$. In the limit $n \to 0$, however, this equation has no solution, and thus one recovers the spectrum of an unperturbed GOE matrix. \\

\subsubsection{Change of basis}
We remark that when $q_1=q^2$, the matrix 
of change of basis $\mathcal{B}^0 \to \mathcal{B}^1$ is exactly the same as in the annealed case, \eqref{eq:app:ChangeBas}. Therefore one can write the projection of this matrix onto the basis of $\tau[{\bm \sigma}^1]$ as 
\begin{align}\label{eq:app:BasisChanged_quench}
\begin{split}
    {\bf R}_{0 \to 1}^{-1} \, \frac{\nabla^2 h({\bm \sigma}^0)}{\sqrt{N-1}}\, {\bf R}_{0 \to 1}\bigg|_\perp&=
    \Pi_q^1
   \;\cdot\frac{\tilde{\mathcal{H}}_Q({\bm \sigma}^0)}{\sqrt{N-1}}\cdot\;
    \Pi_q^1+l_{N-1\, N-1}^0{\bf e}_{N-1}^1 [{\bf e}_{N-1}^1]^T
    \end{split}
\end{align}
where $l_{N-1\, N-1}^0$ is the same as in \eqref{eq:app:SpecialComp}. The other change of basis is obtained in an analogous way.\\

\section{The $p=3$ energy profile: deformation along softest Hessian modes}\label{app:Manipul}
In this Appendix, we determine the expression of the profile \eqref{eq:app:energy_profile} under the assumption that the unconstrained Hessians have the structure and the statistics discussed above, which follows from conditioning on the properties of the various replicas ${\bm \sigma}^a$. As before, we assume $p=3$. We consider both the case in which ${\bf v}$ is chosen to be aligned along the direction of the softest mode of the Hessian $\mathcal{H}({\bm \sigma}^0)$, and the one in which ${\bf v}$ is chosen to be aligned along the direction of the softest mode of the Hessian $\mathcal{H}({\bm \sigma}^1)$.

\subsection{Softest mode at ${\bm \sigma}^0$}
We consider the case in which ${\bf v}$ is chosen to be aligned along the direction of the softest mode ${\bf e}^0_{\rm min}$ of the Hessian $\mathcal{H}({\bm \sigma}^0)$; we introduce the following notation for the minimal eigenvalue of the shifted Hessian, and for the component squared of the corresponding eigenvector along the direction ${\bf e}_{N-1}^0$: 
\begin{equation}
 \frac{\tilde{\mathcal{H}}({\bm \sigma}^0)}{\sqrt{N-1}} \cdot {\bf e}_{\rm min}^0= {\lambda_{\rm min}^0 } \, {\bf e}_{\rm min}^0,    \quad \quad  u^0= ({\bf e}_{\rm min}^0 \cdot  {\bf e}_{N-1}^0)^2.
\end{equation}
We recall from Eq.~\eqref{eq:v_start} the choice:
\begin{align*}
   {\bf v} \to {\bf v}_{\rm soft}^0= \frac{{\bf e}_{\rm min}^0 - \sqrt{  u^0}\, {\bf e}_{N-1}^0}{\sqrt{1-  u^0}}.
\end{align*}
Plugging this expression, we find 
\begin{equation}
\begin{split}
&{\bf v}_{\rm soft}^0 \cdot   \frac{\nabla^2 h({\bm \sigma}^0)}{\sqrt {N-1}} \cdot {\bm \sigma}^1= \tonde{\frac{{\bf e}_{\rm min}^0 - \sqrt{  u^0}\, {\bf e}_{N-1}^0}{\sqrt{1-  u^0}}} \cdot \frac{\nabla^2 h({\bm \sigma}^0)}{\sqrt {N-1}} \cdot \tonde{q  {\bm \sigma}^0-\sqrt{1-q^2} {\bf e}_{N-1}^0}=\\
&
\sqrt{\frac{1-q^2}{1-  u^0}} \tonde{\sqrt{  u^0} \, {\bf e}_{N-1}^0 -{\bf e}_{\rm min}^0 }\, \cdot  \frac{\nabla^2 h({\bm \sigma}^0)}{\sqrt {N-1}} \cdot {\bf e}_{N-1}^0 =\sqrt{\frac{  u^0(1-q^2)}{1-  u^0}}\, \tonde{\mu_0- \lambda_{\rm min}^0},
\end{split}
\end{equation}
and 
\begin{equation}
\begin{split}
{\bf v}_{\rm soft}^0 \cdot   \frac{\nabla^2 h({\bm \sigma}^0)}{\sqrt {N-1}} \cdot {\bf v}_{\rm soft}^0&=\frac{1}{1-  u^0}\tonde{\lambda_{\rm min}^0 +   u^0 \, \mu_0- 2 {  u^0} \lambda_{\rm min}^0}.
\end{split}
\end{equation}
Using a variant of \eqref{eq:app:BasisChanged_annealed} and \eqref{eq:app:BasisChanged_quench}  applied to $\nabla^2h({\bm\sigma^1})$, and the fact that ${\bf v}_{\rm soft}^0\perp {\bf e}_{N-1}^0$, one also obtains 
\begin{equation}\label{eq:app:Third}
\begin{split}
{\bf v}_{\rm soft}^0 \cdot   \frac{\nabla^2 h({\bm \sigma}^1)}{\sqrt {N-1}} \cdot {\bf v}_{\rm soft}^0&=\tonde{\frac{{\bf e}_{\rm min}^0 - \sqrt{  u^0}\, {\bf e}_{N-1}^0}{\sqrt{1-  u^0}}} \cdot \frac{\tilde{\mathcal{H}}_x^1}{{\sqrt {N-1}}} \cdot \tonde{\frac{{\bf e}_{\rm min}^0 - \sqrt{  u^0}\, {\bf e}_{N-1}^0}{\sqrt{1-  u^0}}}\\
&=\frac{1}{1-  u^0} \tonde{{\bf e}_{\rm min}^0 \cdot \frac{\tilde{\mathcal{H}}_x^1}{{\sqrt {N-1}}}  \cdot {\bf e}_{\rm min}^0 - 2 \sqrt{  u^0}\, {\bf e}_{N-1}^0 \cdot \frac{\tilde{\mathcal{H}}_x^1}{{\sqrt {N-1}}}\cdot {\bf e}_{\rm min}^0 + u^0\,\mu_1}\\
&=\frac{1}{1-  u^0} \tonde{{\bf e}_{\rm min}^0 \cdot \frac{\tilde{\mathcal{H}}_x^1}{{\sqrt {N-1}}}  \cdot {\bf e}_{\rm min}^0 - 2 \sqrt{  u^0}\,\sqrt{1-  u^0} \, {\bf v}^0_{\rm soft} \cdot {\bf m}^1 -u^0\,\mu_1},
\end{split}
\end{equation}
where $x \in \{A, Q\}$ and ${\bf m}^1=(m^1_{1 N-1}, \cdots, m^1_{M-1 \, N-1})$.
Now, the definition of ${\bf v}^0_{\rm soft}$ is such that ${\bf e}_{\rm min}^0=\sqrt{1-  u^0}\,{\bf v}^0_{\rm soft}+ \sqrt{  u^0}{\bf e}_{N-1}^0 $, which  implies
\begin{equation}
 \lambda^0_{\rm min} \sqrt{  u^0}=   {\bf e}_{N-1}^0 \cdot \frac{\tilde{\mathcal{H}}_x^0}{{\sqrt {N-1}}}  \cdot {\bf e}_{\rm min}^0 = \sqrt{1-  u^0}\, {\bf v}^0_{\rm soft} \cdot {\bf m}^0+ \sqrt{  u^0}\, \mu_0,
\end{equation}
where $ {\bf m}^0=(m^0_{1 N-1}, \cdots, m^0_{M-1 \, N-1})$. For $p=3$, making use of the fact that ${\bf m}^0={\bf m}^1\equiv {\bf m}$, one finds therefore $ \sqrt{1-  u^0} \, {\bf v}^0_{\rm soft} \cdot {\bf m} = \sqrt{  u^0} (\lambda^0_{\rm min}- \mu_0)$, which plugged into \eqref{eq:app:Third} gives:
\begin{equation}
\begin{split}
{\bf v}_{\rm soft}^0 \cdot   \frac{\nabla^2 h({\bm \sigma}^1)}{\sqrt {N-1}} \cdot {\bf v}_{\rm soft}^0
&=\frac{1}{1-  u^0} \tonde{{\bf e}_{\rm min}^0 \cdot \frac{\tilde{\mathcal{H}}_x^1}{{\sqrt {N-1}}}  \cdot {\bf e}_{\rm min}^0 - 2   u^0\, (\lambda^0_{\rm min}- \mu_0)-u^0\,\mu_1}.
\end{split}
\end{equation}
Therefore, when ${\bf v} \to  {\bf v}_{\rm soft}^0$, the energy profile \eqref{eq:app:energy_profile}, under the conditions that ${\bm \sigma}^a$ are stationary points with energy density $\epsilon_a$, becomes equal to: 
\begin{align}
\label{eq:app:energy_profile_soft0}
\begin{split}
  &     h^{(x)}[\gamma,f] \longrightarrow
\tonde{\gamma^3 +3\gamma^2 \beta \,q} \sqrt{2N}\epsilon_1+ \tonde{\beta^3+3 \beta^2 \gamma \,q} \sqrt{2 N} \epsilon_0+ f^3 h({\bf v}_{\rm soft}^0)+ \gamma \beta f  \sqrt{N-1}\; \sqrt{\frac{  u^0(1-q^2)}{1-  u^0}}\, \tonde{\mu_0- \lambda_{\rm min}^0}\\
& +   \frac{f^2}{2} \sqrt{N-1} \quadre{   \frac{\gamma}{1-  u^0} \tonde{{\bf e}_{\rm min}^0 \cdot \frac{\tilde{\mathcal{H}}_x^1}{{\sqrt {N-1}}}  \cdot {\bf e}_{\rm min}^0 - 2   u^0\, (\lambda^0_{\rm min}- \mu_0)-u^0\,\mu_1} +  \frac{\beta}{1-  u^0}\tonde{\lambda_{\rm min}^0 +   u^0 \, \mu_0- 2 {  u^0} \lambda_{\rm min}^0}}.
\end{split}
\end{align}
This expression depends on four random variables: the minimal eigenvalue $\lambda^0_{\rm min}$ and its eigenvector component $  u^0$, the rescaled energy $h({\bf v}_{\rm soft}^0)$, and the quantity:
\begin{align}\label{eq:app:Chi1}
    \begin{split}
    \chi^0_x:={\bf e}_\text{min}^0 \cdot \frac{\tilde{\mathcal{H}}_x^1}{{\sqrt {N-1}}}  \cdot {\bf e}_\text{min}^0
    \end{split}
\end{align}
where the subscript $x \in \{ Q, A \}$ keeps track of whether we are performing the conditioning of the Hessian statistics under the annealed or the quenched assumptions. The average energy profile is obtained plugging \eqref{eq:app:energy_profile_soft0} into \eqref{eq:app:QuenchedProf} and \eqref{eq:app:Annealed2} and performing the corresponding averages. In the limit of large $N$, however, the quantities $\lambda^0_{\rm min}, u^0$ and $\chi^0$ concentrate, meaning that their distribution converges around their typical value \cite{ros2020distribution, bun2018overlaps}. In this limit, one can therefore replace the quantities in \eqref{eq:app:energy_profile_soft0} with their typical value (under the conditional distribution), and check that the remaining terms in the average factorize with the denominators in the annealed case, or converge to unity in the limit $n \to 0$ in the quenched case. Let us denote with $\lambda^0_{\rm typ}, u^0_{\rm typ}$ and $ \chi^0_{x, \rm typ}$ the typical values of the quantities, and use that the typical value of $h({\bf v}_{\rm soft}^0)$ is equal to zero. \footnote{This follows from the fact that the vector ${\bf v}^0_{\rm soft}$ has zero overlap with ${\bf \sigma}^a$, and thus the random variable  $h({\bf v}_{\rm soft}^0)$ is uncorrelated to any of the random variables on which we are conditioning in this calculation. As a consequence, $h({\bf v}_{\rm soft}^0)$ fluctuates around its average value, that is zero.} Then we find that the energy profile when ${\bf v} \to  {\bf v}_{\rm soft}^0$ reduces to: 
\begin{equation}\label{eq:app:ProfileStart}
\boxed{
\begin{split}
&\epsilon^{(x)}_{{{\bf v}_{\rm soft}^0}}[\gamma, f]:=\lim_{N \to \infty} \frac{h^{(x)}[\gamma, f]}{\sqrt{2N}}=\tonde{\gamma^3 +3\gamma^2 \beta \,q}  \epsilon_1+ \tonde{\beta^3+3 \beta^2 \gamma \,q} \epsilon_0+ \gamma \beta f\; \sqrt{\frac{u^0_{\rm typ}(1-q^2)}{2(1-u^0_{\rm typ})}}\, \tonde{\mu_0- \lambda^0_{\rm typ}}+\\
&\frac{f^2}{2\sqrt{2}} \quadre{\frac{\beta}{1-u^0_{\rm typ}}\tonde{\lambda^0_{\rm typ} + u^0_{\rm typ} \, \mu_0- 2 {u^0_{\rm typ}} \lambda^0_{\rm typ}}- \frac{\gamma}{1-u^0_{\rm typ}} \tonde{ u^0_{\rm typ}\,\mu_1+2 u^0_{\rm typ}\, (\lambda^0_{\rm typ}- \mu_0)}}+
\frac{f^2}{2\sqrt{2}} \frac{\gamma}{1-u^0_{\rm typ}} \;  \chi^0_{x, \rm typ} 
\end{split}}
\end{equation}
$x \in\{Q, A\}$  , where we have 
\begin{equation}\label{eq:app:OvTyp}
\begin{split}
   \chi^0_{Q, \rm typ}&= \lim_{n \to 0} \mathbb{E}^{\rm cond}_n\quadre{{\bf e}_{\rm min}^0 \cdot \frac{\tilde{\mathcal{H}}_Q^1}{{\sqrt {N-1}}}  \cdot {\bf e}_{\rm min}^0},\\
   \chi^0_{A, \rm typ}&= \mathbb{E}^{\rm cond}_1\quadre{{\bf e}_{\rm min}^0 \cdot \frac{\tilde{\mathcal{H}}_A^1}{{\sqrt {N-1}}}  \cdot {\bf e}_{\rm min}^0}.
   \end{split}
\end{equation}

The typical values $\lambda^0_{\rm typ}, u^0_{\rm typ}$ are known, and discussed in Appendix \ref{app:statistics_HESSIANS}. To get the energy profile, it remains to determine the quantities \eqref{eq:app:OvTyp}. We discuss this in Appendix \ref{app:Overlaps}.

\subsection{Softest mode at ${\bm \sigma}^1$}
Let us now consider the case in which ${\bf v}$ is chosen to be aligned along the direction of the softest mode of $\mathcal{H}({\bm \sigma}^1)$, i.e., in the same notation as above, 
\begin{align*}
   {\bf v} \to {\bf v}_{\rm soft}^1= \frac{{\bf e}_{\rm min}^1 - \sqrt{  u^1}\, {\bf e}_{N-1}^1}{\sqrt{1-  u^1}},\quad  \quad    u^1= ({\bf e}_{\rm min}^1 \cdot  {\bf e}_{N-1}^1)^2, \quad \quad \frac{\tilde{\mathcal{H}}({\bm \sigma}^1)}{\sqrt{N-1}} \cdot {\bf e}_{\rm min}^1= {\lambda_{\rm min}^1 } {\bf e}_{\rm min}^1.
\end{align*}
In this case, using that for $p=3$ one has $\nabla^2 h({\bm \sigma}^0) \cdot  {\bm \sigma}^1=\nabla^2 h({\bm \sigma}^1) \cdot  {\bm \sigma}^0$, we find
\begin{equation}\label{eq:sotto}
\begin{split}
{\bf v}_{\rm soft}^1 \cdot   \frac{\nabla^2 h({\bm \sigma}^0)}{\sqrt{N-1}} \cdot {\bm \sigma}^1&={\bf v}_{\rm soft}^1 \cdot   \frac{\nabla^2 h({\bm \sigma}^1)}{\sqrt {N-1}} \cdot {\bm \sigma}^0= \tonde{\frac{{\bf e}_{\rm min}^1 - \sqrt{  u^1}\, {\bf e}_{N-1}^1}{\sqrt{1-  u^1}}} \cdot \frac{\nabla^2 h({\bm \sigma}^1)}{\sqrt {N-1}} \cdot \tonde{q  {\bm \sigma}^1-\sqrt{1-q^2} {\bf e}_{N-1}^1}\\
&=
\sqrt{\frac{1-q^2}{1-  u^1}} \tonde{ \sqrt{  u^1}\,  {\bf e}_{N-1}^1 -{\bf e}_{\rm min}^1 }\, \cdot  \frac{\nabla^2 h({\bm \sigma}^1) }{\sqrt {N-1}} \cdot {\bf e}_{N-1}^1 \\
&=\sqrt{\frac{  u^1(1-q^2)}{1-  u^1}}\, \tonde{\mu_1- \lambda_{\rm min}^1},
\end{split}
\end{equation}
and 
\begin{equation}
\begin{split}
{\bf v}_{\rm soft}^1 \cdot   \frac{\nabla^2 h({\bm \sigma}^1)}{\sqrt {N-1}} \cdot {\bf v}_{\rm soft}^1&=\frac{1}{1-  u^1}\tonde{\lambda_{\rm min}^1 +   u^1 \, \mu_1- 2 {  u^1} \, \lambda_{\rm min}^1}
\end{split}
\end{equation}
and finally 
\begin{equation}
\begin{split}
{\bf v}_{\rm soft}^1 \cdot   \frac{\nabla^2 h({\bm \sigma}^0)}{\sqrt {N-1}} \cdot {\bf v}_{\rm soft}^1&=\frac{1}{1-  u^1} \tonde{{\bf e}_{\rm min}^1 \cdot \frac{\tilde{\mathcal{H}}_x^0}{{\sqrt M}}  \cdot {\bf e}_{\rm min}^1 - 2   u^1\, (\lambda^1_{\rm min}- \mu_1)-u^1\,\mu_0}.
\end{split}
\end{equation}
Therefore, when ${\bf v} \to  {\bf v}_{\rm soft}^1$, the energy profile \eqref{eq:app:energy_profile}, under the conditions that ${\bm \sigma}^a$ are stationary points with energy density $\epsilon_a$, becomes equal to \eqref{eq:app:energy_profile_soft0} by swapping the indices $0$ and $1$, and same for $\beta$ and $\gamma$, in the terms that contain $f$. Introducing
\begin{align}\label{eq:app:Chi0}
    \begin{split}
       \chi^1_x:= {\bf e}_\text{min}^1 \cdot \frac{\tilde{\mathcal{H}}_x^0}{{\sqrt {N-1}}}  \cdot {\bf e}_\text{min}^1,
    \end{split}
\end{align}
as well as the typical values as before, we get that the energy profile in this case reads:
\begin{equation}
\label{eq:app:ProfileEnd}
\boxed{
\begin{split}
&\epsilon^{(x)}_{ {\bf v}_{\rm soft}^1}[\gamma, f]:=\lim_{N \to \infty} \frac{h^{(x)}[\gamma, f]}{\sqrt{2N}}=\tonde{\gamma^3 +3\gamma^2 \beta \,q}  \epsilon_1+ \tonde{\beta^3+3 \beta^2 \gamma \,q} \epsilon_0+ \gamma \beta f \; \sqrt{\frac{u^1_{\rm typ}(1-q^2)}{2(1-u^1_{\rm typ})}}\, \tonde{\mu_1- \lambda^1_{\rm typ}}+\\
&\frac{f^2}{2\sqrt{2}} \quadre{\frac{\gamma}{1-u^1_{\rm typ}}\tonde{\lambda^1_{\rm typ} + u^1_{\rm typ} \, \mu_1- 2 {u^1_{\rm typ}} \lambda^1_{\rm typ}}- \frac{\beta}{1-u^1_{\rm typ}} \tonde{u^1_{\rm typ}\,\mu_0+ 2 u^1_{\rm typ}\, (\lambda^1_{\rm typ}- \mu_1)}}+
\frac{f^2}{2\sqrt{2}} \frac{\beta}{1-u^1_{\rm typ}} \;  \chi^1_{x, \rm typ}
\end{split}}
\end{equation}
$x \in\{Q, A\}$,  where now
\begin{equation}\label{eq:app:OvTyp1}
\begin{split}
   \chi^1_{Q, \rm typ}&= \lim_{n \to 0} \mathbb{E}^{\rm cond}_n\quadre{{\bf e}_{\rm min}^1 \cdot \frac{\tilde{\mathcal{H}}_Q^0}{{\sqrt {N-1}}}  \cdot {\bf e}_{\rm min}^1},\\
   \chi^1_{A, \rm typ}&= \mathbb{E}^{\rm cond}_1\quadre{{\bf e}_{\rm min}^1 \cdot \frac{\tilde{\mathcal{H}}_A^0}{{\sqrt {N-1}}}  \cdot {\bf e}_{\rm min}^1}.
   \end{split}
\end{equation}

Notice that despite the analogy of the expression, in the quenched setting the problem is not symmetric since the statistics of the Hessian matrices is different for $a=0$ and $a=1$, and therefore the typical values of $\lambda_{\rm min}^a, u^a$ can be quite different for $a=0$ and $a=1$, as we commented in Appendix \ref{app:statistics_HESSIANS}. We discuss the quantities \eqref{eq:app:OvTyp1} in Appendix~\ref{app:Overlaps}. \\

\section{Eigenvectors overlaps}\label{app:Overlaps}

The manipulations of Appendix \ref{app:Manipul} show that in order to determine the typical value of the energy profile, one has to compute the typical value of the terms \eqref{eq:app:Chi1} and \eqref{eq:app:Chi0}. 
Using the eigenvalue decomposition $\tilde{\mathcal{H}}^a_x=\sum_{\alpha=1}^{N-1}\tilde \lambda_\alpha^a{\bf u}_\alpha^a[{\bf u}_\alpha^a]^T$, for $x\in\{A,Q\}$ and $a\in\{0,1\}$, we can formally write
\begin{align}
    \begin{split}
       \chi^0_x= \sum_{\alpha=1}^{N-1}\tilde \lambda_\alpha^1 \,[ {\bf e}_\text{min}({\bm \sigma}^0) \cdot {\bf u}_\alpha^1]^2, \quad \quad \chi^1_x= \sum_{\alpha=1}^{N-1}\tilde \lambda_\alpha^0 \,[ {\bf e}_\text{min}({\bm \sigma}^1) \cdot {\bf u}_\alpha^0]^2,
    \end{split}
\end{align}
where the dependence of the eigenvalue and eigenvector distribution on $x$ is implicit. In the limit of large matrix size $N\gg 1$, the quantities $\chi^a_x$ with $a=0,1$ converge to their average.  Following Refs.~\cite{bun2018overlaps, paccoros}, we introduce the averaged squared overlap between eigenstates of two different, correlated matrices of size $(N-1) \times (N-1)$:
\begin{equation}\label{eq:app:OverlapDef}
    \Phi(\lambda^0_\alpha,  \lambda^1_\beta):= (N-1) \mathbb{E} [ ({\bf u}_{\lambda^0_\alpha} \cdot {\bf u}_{\lambda^{1}_\beta})^2],
\end{equation}
where the expectation is over the realizations of the random matrices. We remark that such expression can take different expressions whether there are zero, one or two isolated eigenvalues among $\lambda^0_\alpha,\lambda^1_\beta$ \cite{paccoros}. We define with $\tilde{\rho}_N^a(\lambda)$ the average eigenvalues density of the matrices $\tilde{\mathcal{H}}^a_x$ (the leading order term, as well as eventual subleading corrections corresponding to isolated eigenvalues of the matrix). Then the term to be determined can be formally written as:
\begin{equation}
    \chi^0_{x, \rm typ}= \int \lambda \,\tilde{\rho}_N^1(\lambda)\,    \Phi(\lambda_{\rm \min}^0, \lambda)\,d \lambda, \quad \quad \chi^1_{x, \rm typ}= \int   \lambda\,\tilde{\rho}_N^0(\lambda) \,   \Phi(\lambda_{\rm \min}^1, \lambda)\,d \lambda.
\end{equation}
In order to compute these terms, one has to determine both the density $\tilde{\rho}_N^a(\lambda)$ of each matrix  $\tilde{\mathcal{H}}^a_x$  (paying attention to the possible existence of isolated eigenvalues), as well as the typical overlaps between the eigenvectors of $\tilde{\mathcal{H}}^a_x$ and the minimal eigenvector of $\tilde{\mathcal{H}}^{b \neq a}_x$, which are correlated according to  Appendix \ref{app:statistics_HESSIANS}. In the following, we discuss the quenched and annealed case separately.\\\\

\subsection{Eigenvectors overlaps: the annealed setting. }
Without loss of generality, we focus on the case in which ${\bf v} \to {\bf v}_{\rm soft}^0$, and we have to determine the typical value
\begin{equation}
    \chi^0_{A, \rm typ}=  \mathbb{E}^{\rm cond}_1\quadre{\; {\bf e}_{\rm min}^0 \cdot \frac{\tilde{\mathcal{H}}_A^1}{{\sqrt {N-1}}}  \cdot {\bf e}_{\rm min}^0\; },
\end{equation}
where ${\bf e}_{\rm min}^0$ is the eigenvector associated to the minimal eigenvalue of the shifted Riemannian Hessian $\tilde 
{\mathcal{H}}_A({\bm \sigma}^0)$ given in \eqref{eq:app:TildeMatDef}. As we argued in Appendix~\ref{app:AnnealedDistribution}, the matrix  $\tilde{\mathcal{H}}^1_A$ can have one or zero isolated eigenvalues. On the contrary, while $\tilde{\mathcal{H}}^0_A$ could a priori present an isolated eigenvalue, it can be checked that for no value of $\epsilon_{\rm gs}\leq \epsilon_0,\epsilon_1\leq \epsilon_{\rm th}$ and $q$, does $\mu_0$ generate an isolated eigenvalue in the region of parameters in which the complexity $\Sigma(\epsilon_1,q|\epsilon_0)>0$. Therefore, the minimal eigenvalue is at the edge of the semicircle,  $\lambda_{\rm min}^0=-2\sigma$. This implies that the rescaled overlap \eqref{eq:app:OverlapDef} is of $\mathcal{O}(1)$ for all possible values of $\lambda$. \cite{bun2018overlaps, paccoros}.
In the limit of $N\to+\infty$ we have 
\begin{align}
&\tilde\rho_N^1(\lambda)=\rho_\sigma(\lambda)+\frac{\Theta\left(|\mu_1|-2\sigma+\Delta^2/\sigma \right)}{N}\delta(\lambda-\lambda^1_{\rm iso})+ \mathcal{O}\tonde{\frac{1}{N^2}}, \quad \quad \rho_\sigma(\lambda)=\frac{\sqrt{4\sigma^2-\lambda^2}}{2\pi\sigma^2},
\end{align}
where $\Theta$ is non-zero only whenever \eqref{eq:app:ConditionEx} holds. Then we obtain 
\begin{equation}
        \chi^0_{\rm typ, A}= \int\lambda \,\rho_\sigma(\lambda)\,   \Phi(\lambda_{\rm \min}^0, \lambda) d \lambda+ \frac{\Theta}{N}\lambda_{\rm iso}^1\Phi(\lambda_{\rm min}^0,\lambda_{\rm iso}^1)+ \mathcal{O}\tonde{\frac{1}{N^2}}=\int \lambda\,\rho_\sigma(\lambda)\, \lambda\,    \Phi(\lambda_{\rm \min}^0, \lambda)d \lambda + \mathcal{O}\tonde{\frac{1}{N}}.
\end{equation}
From Refs.~\cite{bun2018overlaps, paccoros} one can get the expression for the rescaled, squared overlaps $\Phi(\lambda_{\rm \min}^0, \lambda)$ between eigenvalues belonging to the  continuous part of the distribution. In the specific case in which $\lambda_{\rm \min}^0=-2 \sigma$, this reads
\begin{align}
\Phi(-2 \sigma, \lambda)= 
\frac{\sigma^4 \sigma_W^2 (2 \sigma^2 - \sigma_W^2)}{\left(2 \sigma^4 - 2 \sigma^2 \sigma_W^2 + \sigma_W^4 + \sigma^3 \lambda - 
  \sigma \sigma_W^2 \lambda\right)^2}+ \mathcal{O}\tonde{\frac{1}{N}},\quad\quad\lambda\in[-2\sigma,2\sigma]
\end{align}
where the parameters given in \eqref{eq:app:FormPar} for general values of $p$.  Therefore
\begin{align}
  \chi^0_{A, \rm typ}&= \int_{-2\sigma}^{2\sigma} \lambda\,\frac{\sqrt{4 \sigma^2 - \lambda^2}}{2 \pi } \,  \frac{\sigma^2 \sigma_W^2 (2 \sigma^2 - \sigma_W^2)}{\left(2 \sigma^4 - 2 \sigma^2 \sigma_W^2 + \sigma_W^4 + \sigma^3 \lambda - 
  \sigma \sigma_W^2 \lambda\right)^2}d \lambda   + \mathcal{O}\tonde{\frac{1}{N}}\\
  &=\frac{1}{2}\sigma^2\sigma_W^2(2\sigma^2-\sigma_W^2)\frac{a}{b^3}\frac{c^2+2\sqrt{1-c^2}-2}{\sqrt{1-c^2}}    + \mathcal{O}\tonde{\frac{1}{N}}
\end{align}
where 
\begin{equation}
\begin{cases}
    a=2\sigma^4-2\sigma^2\sigma_W^2+\sigma_W^4\\
    b=\sigma^3-\sigma\sigma_W^2\\
    c=2\sigma \,b/a.
\end{cases}
\end{equation}
For $p=3$ one has $\sigma=\sqrt{6}$ and $\sigma_W=\sqrt{6 (1-q)}$ and this expression reduces to:
\begin{equation}\label{eq:SimplifiedIntgral}
    \chi^0_{A, \rm typ}=\frac{1}{2}\sigma^2\sigma_W^2(2\sigma^2-\sigma_W^2)\frac{a}{b^3}\frac{c^2+2\sqrt{1-c^2}-2}{\sqrt{1-c^2}}=-2\sqrt{6}\,q.
\end{equation}
\\

Let us now consider $\chi^1_{A, \rm typ}$. 
As we have remarked above, there exist values of $\epsilon_0,\epsilon_1,q$ such that $\Sigma(\epsilon_1,q|\epsilon_0)>0$ and such that the matrix $\tilde{\mathcal{H}}^1_A$ exhibits an isolated eigenvalue (meaning that $\mu_1$ satisfies Eq.~\eqref{eq:app:ConditionEx}). It can be checked that for this regime of parameters $\mu_1$ is negative, and so it is the isolated eigenvalue, given that the two quantities have the same sign \cite{paccoros}. Therefore, when the isolated eigenvalue exists, it is the minimal eigenvalue of $\tilde{\mathcal{H}}^1_A$.
On the other hand, we find that $\mu_0$ does not generate an isolated eigenvalue in the regime of parameters we are interested in. When $\lambda_{\rm min}^1$ is isolated, i.e. when $\mu_1$ satisfies Eq.~\eqref{eq:app:ConditionEx}, to compute $\chi^1_{A, \rm typ}$ we have use a different expression for the rescaled, squared overlap $\Phi$. In general, we can write 
\begin{align}
  \chi^1_{A, \rm typ}&= \int_{-2\sigma}^{2\sigma} \lambda\,\frac{\sqrt{4 \sigma^2 - \lambda^2}}{2 \pi\sigma^2 }\,\left[\Phi(-2\sigma,\lambda)\left(1-\Theta\left(|\mu_1|-2\sigma+\Delta^2/\sigma \right)+\Phi(\lambda^1_{\rm iso},\lambda)\Theta\left(|\mu_1|-2\sigma+\Delta^2/\sigma \right)\right)
  \right]\,d \lambda + \mathcal{O}\tonde{\frac{1}{N}}
\end{align}
where the expression for $\Phi(\lambda^1_{\rm iso},\lambda)$ has been computed in Ref.~\cite{paccoros}, and applied to our case reads
\begin{align}
\label{eq:app:Phi_iso_bulk}
\begin{split}
\Phi(\lambda^1_{\rm iso},y)&=\mathfrak{q}_{\sigma, \Delta}(\lambda_{\rm iso}^1,\mu_1)\Bigg[\frac{4\Delta^2\sigma^4}{\sqrt{[\lambda_{\rm iso}^1]^2-4\sigma^2}}\frac{bc-ad}{c^2+d^2}-4\sigma^4\Delta^4\frac{b_1c_1e_1-a_1d_1e_1-a_1c_1f_1-b_1d_1f_1(4\sigma^2-y^2)}{(c_1^2+d_1^2)(e_1^2+f_1^2)}\\
&-8\sigma^4\Delta^2\frac{c_1b_2-a_2b_2\Delta^2-a_2c_1\sigma_W^2-\sigma_W^2\Delta^2(4\sigma^2-y^2)}{(c_1^2+\Delta^4)(b_2^2+\sigma_W^4)}+\frac{\Delta^4\mathfrak{g}_\sigma(\lambda_{\rm iso}^1)}{(\lambda_{\rm iso}^1-\mu_1)}\frac{1}{a_3^2+b_3^2}
\Bigg]
\end{split}
\end{align}
with parameters
\begin{equation}
    \begin{split}
&a=-(4\sigma^2-\lambda_{\rm iso}^1 y)\\
&b=\sqrt{[\lambda_{\rm iso}^1]^2-4\sigma^2}\\
&c=(2\sigma^2 - \sigma_W^2)^2 (\lambda_{\rm iso}^1 - y)^2 - 
 2\sigma_W^2 (2 \sigma^2 - \sigma_W^2) \sqrt{  [\lambda_{\rm iso}^1]^2-4 \sigma^2} \;(\lambda_{\rm iso}^1 - y)  + 
 \sigma_W^4 ( [\lambda_{\rm iso}^1]^2-4 \sigma^2 ) + \sigma_W^4 (y^2 -4 \sigma^2 )\\
&d=-2\sigma_W^2\;
   \left[-\sigma_W^2\sqrt{ [\lambda_{\rm iso}^1]^2 -4 \sigma^2}
      + (2\sigma^2 - \sigma_W^2) (\lambda_{\rm iso}^1 - y)\right]\\
&a_1=(\lambda_{\rm iso}^1 - y)^2 - 2\sqrt{ [\lambda_{\rm iso}^1]^2-4\sigma^2 } \,(y-\lambda_{\rm iso}^1 ) + [\lambda_{\rm iso}^1]^2 +  y^2-8\sigma^2\\
&b_1=-2 \; \left[y-\lambda_{\rm iso}^1  - \sqrt{[\lambda_{\rm iso}^1]^2 -4 \sigma^2 }\right]\\
&c_1=-2\mu_0\sigma^2-\Delta^2y+2\sigma^2y\\
&d_1=\Delta^2\\
&e_1=(\sigma_W^2-2\sigma^2 )^2 (\lambda_{\rm iso}^1 - y)^2 - 
 2\sigma_W^2 (2\sigma^2 - \sigma_W^2)\sqrt{[\lambda_{\rm iso}^1]^2-4 \sigma^2 }\, (\lambda_{\rm iso}^1- y) + \sigma_W^4 ([\lambda_{\rm iso}^1]^2-4 \sigma^2 ) + \sigma_W^4 ( y^2-4 \sigma^2 )\\ 
&f_1=-2\sigma_W^2  \left[(2\sigma^2 - \sigma_W^2) (\lambda_{\rm iso}^1 - y) - \sigma_W^2\sqrt{[\lambda_{\rm iso}^1]^2-4 \sigma^2}\,\right]\\
&a_2=\lambda_{\rm iso}^1-y+\sqrt{[\lambda_{\rm iso}^1]^2-4\sigma^2}\\
&b_2=(\sigma_W^2-2 \sigma^2 ) (\lambda_{\rm iso}^1 - y) + \sigma_W^2 \sqrt{[\lambda_{\rm iso}^1]^2-4 \sigma^2}\\
&a_3=y-\mu_0 - \frac{\Delta^2 y}{2 \sigma^2}\\
&b_3=\frac{\Delta^2}{2 \sigma^2},
\end{split}
\end{equation}
and where the function $\mathfrak{q}_{\sigma, \Delta}$ and the explicit expression of $\lambda_{\rm iso}^1$ are given in \eqref{eq:Qfunc} and \eqref{eq:app:IsoExplicit}.\\

\subsection{Eigenvectors overlaps: the quenched setting. }
 We discuss first the case in which  ${\bf v} \to {\bf v}_{\rm soft}^0$, and we have to determine the expectation 
\begin{equation}
    \chi^0_{Q, \rm typ}= \lim_{n \to 0}   \mathbb{E}^{\rm cond}_n\quadre{\; {\bf e}_{\rm min}^0 \cdot \frac{\tilde{\mathcal{H}}_Q^1}{{\sqrt {N-1}}}  \cdot {\bf e}_{\rm min}^0\; }.
\end{equation}
In this case ${\bf e}_{\rm min}^0$ is the eigenvector associated to the minimal eigenvalue of the shifted Riemannian Hessian $\tilde{\mathcal{H}}_Q^0$, which together with $\tilde{\mathcal{H}}_Q^1$ is defined in Eq.~\eqref{eq:app:quenched_matrix_1} and \eqref{eq:app:Quench00} respectively. As argued in Appendix \ref{app:QuenchedDistribution}, in the limit $n\to 0$, the matrix $\tilde{\mathcal{H}}_Q^0$ has no isolated eigenvalues, because the zeros of equation \eqref{eq:app:egval_eqn_0} become poles in the limit $n \to 0$.
On the contrary, in the limit $n\to 0$ the matrix $\tilde{\mathcal{H}}_Q^1$ could present an isolated eigenvalue depending on the strength of $\mu_1$. Hence, $\lambda_{\rm min}^0=-2\sigma$, and by applying the same reasoning as in the annealed case, we are left with the fact that any possible contribution $\Phi(\lambda_{\rm min}^0,\lambda^1_{\rm iso})$ comes at an order of $1/N$ an can be neglected to leading order at large $N$. Hence, one ends up with 
\begin{align*}
     \chi^0_{Q, \rm typ}= \chi^0_{A, \rm typ}.
\end{align*}

We turn now to the case ${\bf v} \to {\bf v}_{\rm soft}^1$, where we need to determine
\begin{equation}
    \chi^1_{Q,\rm typ}=  \lim_{n \to 0} \mathbb{E}^{\rm cond}_n\quadre{\; {\bf e}_{\rm min}^1 \cdot \frac{\tilde{\mathcal{H}}_Q^0}{{\sqrt {N-1}}}  \cdot {\bf e}_{\rm min}^1\; }.
\end{equation}
This case is a slightly more complicated, since  $\tilde{\mathcal{H}}^1_Q$ can display an isolated eigenvalue in the limit $n \to 0$, and thus one has to consider how the function $\Phi(\lambda^1_{\rm iso}, \lambda)$ changes in the quenched case, for $\lambda\in[-2\sigma,2\sigma]$. The way to proceed is to replicate the computation in Ref.~\cite{paccoros} in the case of finite $n$, and then send $n\to 0$. It turns out that this calculation reproduces the result in Eq.~\eqref{eq:app:Phi_iso_bulk}, implying that
\begin{align*}
    \chi^1_{Q,\rm typ}=\chi^1_{A,\rm typ}.
\end{align*}
\\

\subsection{ On quenched \emph{vs} annealed. }
As we recalled in Appendix \ref{app:recap_paper_saddles}, when computing the complexity $\Sigma(\epsilon_1, q |\epsilon_0)$ one obtains the identity \eqref{eq:QuenchedEqualAnnealed}, which implies that the calculation performed within the quenched formalism (making use of the replica trick with $n \to 0$) gives the same result as that performed within  the annealed formalism (which corresponds to $n=1$). This follows from the fact that the typical value of the overlap parameter $q_1$, giving the typical overlap between the replicas of the secondary configuration ${\bm \sigma}^1$, takes the particularly simple value $q_1=q^2$. This identity does not imply a priori that the typical energy profile computed within the quenched prescription is the same as that computed within the annealed prescription, since the statistics of the conditional Hessians (in particular, the finite rank perturbations) is different in the two cases. We find however that the two prescription give the same result for the energy profile, too, to leading order in $N$. In fact, the main difference between the two prescriptions is in the statistics of the conditional Hessian $\mathcal{H}({\bm\sigma}^0)$: in the quenched case, this Hessian exhibits $n$ special line and columns with identical statistical properties, whose effect vanishes as $n \to 0$. In this limit, one recovers the statistics of the unconditional Hessian. This is consistent with the fact that in the quenched setup the reference configuration ${\bm \sigma}^0$ is selected first, independently of the secondary one ${\bf \sigma}^1$, and therefore the statistics of its Hessian should not be affected by the coupling to ${\bf \sigma}^1$. On the contrary, in the annealed framework the primary and secondary configurations are treated on the same footing, and the conditional distribution of the Hessian at  ${\bm \sigma}^0$ has a special line and column that depends on the correlation with  ${\bm \sigma}^1$. The two prescriptions give different results whenever this special line affects the eigenvalue distribution of the Hessian by generating isolated eigenvalues, which are not there in the quenched case. However, we find that this circumstance is never realized for the values of parameters $\epsilon_a, q$ that are of interest (such that $\Sigma(\epsilon_1, q|\epsilon_0)>0$, $\epsilon_{\rm gs}\leq \epsilon_a\leq\epsilon_{\rm th}$ and $q\in[0,1]$). As a consequence, in the regime of parametrs that we ar interested in the typical energy profile  is the same with both prescriptions. For this reason, we will now drop the $Q$ or $A$ pedices in the following.

\section{Resulting energy profiles }
\label{app:resulting_energy_profiles}
We give in this Appendix the final result for the typical energy profile, in the case $p=3$. We separate the cases in which ${\bf v}$ is chosen to be related to the softest mode of the Hessian at the reference and secondary configuration, respectively. \\

{\bf Softest mode at ${\bm \sigma}^0$. }
As it follows from the discussion of Appendix \ref{app:Manipul} and \ref{app:Overlaps}, in this case $\lambda_{\rm typ}^0=-2 \sigma= - 2 \sqrt{6}$; Moreover, ${\bf e}_{\text{min}}({\bm \sigma}^0)$ is a random vector uniformly distributed on $\mathcal{S}_N(1)$, so that its typical overlap with ${\bf e}_{N-1}({\bm \sigma}^0)$ is $u^0_{\rm typ}={\bf e}_\text{min}({\bm \sigma}^0)\cdot {\bf e}_{N-1}({\bm \sigma}^0)=0$. Using our previously found results, the energy profile \eqref{eq:app:ProfileStart} simplifies into:
\begin{align*}
\begin{split}
&\epsilon_{{\bf v}_{\rm soft}^0}[\gamma, f]=\tonde{\gamma^3 +3\gamma^2 \beta \,q}  \epsilon_1+  \tonde{\beta^3+3 \beta^2 \gamma \,q}\epsilon_0-\sqrt{3}\,f^2(\gamma) \, \beta -f^2(\gamma)\,\gamma\sqrt{3}\,q,
\end{split}
\end{align*}
which corresponds to the Eq.~\eqref{eq:energy_profile_quench_0} in the main text. \\

{\bf Softest mode at ${\bm \sigma}^1$. }
In this case there are two possibilities: if {there is no isolated eigenvalue} at $\tilde{\mathcal{H}}({\bm\sigma}^1)$, then $\lambda^1_\text{min}=-2\sigma$ and we obtain a similar expression as above:
\begin{align*}
\begin{split}
&\epsilon_{{\bf v}_{\rm soft}^1}[\gamma, f]=\tonde{\gamma^3 +3\gamma^2 \beta \,q}  \epsilon_1+  \tonde{\beta^3+3 \beta^2 \gamma \,q}\epsilon_0-\sqrt{3}\,f^2(\gamma) \, \gamma -f^2(\gamma)\,\beta \, \sqrt{3}q\\
\end{split}
\end{align*}

\noindent If { there is a negative isolated eigenvalue}, then the formula changes, and Eq.~\eqref{eq:app:ProfileEnd} becomes:
\begin{align*}
&\epsilon_{ {\bf v}_{\rm soft}^1}[\gamma, f]=\tonde{\gamma^3 +3\gamma^2 \beta \,q}  \epsilon_1+ \tonde{\beta^3+3 \beta^2 \gamma \,q} \epsilon_0+ \gamma \beta f \; \sqrt{\frac{u^1_{\rm typ}(1-q^2)}{2(1-u^1_{\rm typ})}}\, \tonde{\mu_1- \lambda^1_{\rm typ}}+\\
&\frac{f^2}{2\sqrt{2}} \quadre{\frac{\gamma}{1-u^1_{\rm typ}}\tonde{\lambda^1_{\rm typ} + u^1_{\rm typ} \, \mu_1- 2 {u^1_{\rm typ}} \lambda^1_{\rm typ}}- \frac{\beta}{1-u^1_{\rm typ}} \tonde{u^1_{\rm typ}\,\mu_0+ 2 u^1_{\rm typ}\, (\lambda^1_{\rm typ}- \mu_1)}}+
\frac{f^2}{2\sqrt{2}} \frac{\beta}{1-u^1_{\rm typ}} \;  \chi^1_{\rm iso}
\end{align*}
where we recall the values of the various quantities:
\begin{align*}
&\lambda^1_{\rm typ}=\frac{{2 \mu_1 \sigma^2- \Delta^2 \mu_1- \text{sign}(\mu_1) \Delta^2 \sqrt{\mu_a^2-4 (\sigma^2- \Delta^2)}}}{2 (\sigma^2-\Delta^2)}\\
&u^1_{\rm typ}=\mathfrak{q}_{\sigma,\Delta}(\lambda_{\rm typ}^1,\mu_1)\\
&\chi^1_{\rm iso}:=\int_{-2\sigma}^{2\sigma}\lambda\,\frac{\sqrt{4\sigma^2-\lambda^2}}{2\pi\sigma^2}\Phi(\lambda^1_{\rm typ},\lambda)\,d\lambda
\end{align*}
where the integral $\chi^1_{\rm iso}$ can be easily computed numerically. This expression coincides with Eq.~\eqref{eq:energy_quench_1} in the main text. 

\section{Gradient along the geodesic path}
\label{app:gradient_max_barr}
In this Appendix, we derive the statistical distribution of the vector $ \nabla h({\bm\sigma}[\gamma;f=0])$ at each configuration along the geodesic path, parametrized as \eqref{eq:path} with $f(\gamma)=0$. We denote each configuration along the path simply as ${\bm \sigma}(\gamma):= {\bm\sigma}[\gamma;0]$, and take $p=3$. For any fixed value of $\gamma\in(0,1)$,  plugging the expression for ${\bm\sigma}(\gamma)$ inside the formula for $\nabla h$ one gets
\begin{align}
    \nabla h({\bm\sigma}(\gamma))=\gamma^2\,\nabla h({\bm\sigma}^1)+\beta^2\,\nabla h({\bm\sigma}^0)+\gamma\,\beta\,\nabla^2h({\bm\sigma}^0)\cdot {\bm\sigma}^1,
\end{align}
which using \eqref{eq:Idi} and conditioning on the properties of ${\bm \sigma}^a$ reduces to:
\begin{align}
\begin{split}
    \nabla h({\bm\sigma}(\gamma))&=\gamma^2 3 \sqrt{2N}\epsilon_1{\bm\sigma}^1+\beta^2 3 \sqrt{2N}\epsilon_0{\bm\sigma}^0+\gamma\beta\nabla^2 h({\bm\sigma}^0)\cdot \left(q{\bm\sigma}^0-\sqrt{1-q^2}{\bf e}_{N-1}({\bm\sigma}^0)\right)\\
    &=\gamma^2 3 \sqrt{2N}\epsilon_1{\bm\sigma}^1+\beta^2 3\sqrt{2N}\epsilon_0{\bm\sigma}^0+\gamma\beta\left[ 
    6q\, \sqrt{2N}\epsilon_0{\bm\sigma}^0-\sqrt{1-q^2}\,\nabla^2h({\bm\sigma}^0)\cdot {\bf e}_{N-1}({\bm\sigma}^0)
    \right].
\end{split}
\end{align}
We now consider the vector ${\bf g}(\gamma):={\bf g}({\bm \sigma}(\gamma))$, that is the projection of $\nabla h$ on the $(N-1)$-dimensional tangent plane $\tau[{\bm\sigma}(\gamma)]$. We choose a basis of this tangent plane in such a way that the first $N-2$ elements span the subspace orthogonal to both ${\bm\sigma}^0,{\bm\sigma}^1$; we denote them with ${\bf x}_1,\ldots,{\bf x}_{N-2}$ as in Appendix \ref{app:geometrical_context}. The remaining vector equals to
\begin{align}
    {\bf e}_{N-1}(\gamma):= {\bf e}_{N-1}({\bm \sigma}(\gamma))=A({\bm\sigma}^0+C{\bm\sigma}^1), \quad \quad A=(1+2Cq+C^2)^{-1/2}, \quad \quad C=-\frac{\gamma q+\beta}{\gamma +\beta q}.
\end{align}
This vector is orthogonal to  $ {\bm\sigma}(\gamma)$ and has unit norm. It is simple to check that it coincides with the tangent vector to the geodesic path at the point ${\bm \sigma}(\gamma)$. Therefore, the gradient ${\bf g}(\gamma)= {\bf g}^\parallel(\gamma)+ {\bf g}^\perp(\gamma)$, where ${\bf g}^\parallel(\gamma) = ({\bf g}(\gamma) \cdot {\bf e}_{N-1}(\gamma)) {\bf e}_{N-1}(\gamma)$ is the component tangent to the geodesic, while ${\bf g}^\perp(\gamma)$ is the orthogonal one. Using that 
\begin{align}
    {\bf e}_{N-1}(\gamma)=
    (A+ACq){\bm\sigma}^0-AC\sqrt{1-q^2}{\bf e}_{N-1}({\bm\sigma}^0)
\end{align}
 we see that the tangent component equals to:
\begin{align}
    \begin{split}
       \nabla h({\bm\sigma}(\gamma))\cdot {\bf e}_{N-1}(\gamma)&=3\gamma^2  \sqrt{2N}\epsilon_1(A q + A C)+3\beta^2 \sqrt{2N}\epsilon_0(A+ACq)+\gamma\beta\bigg[6q\,\sqrt{2N}\epsilon_0(A+ACq)\\
        &-\sqrt{1-q^2}\,{\bf e}_{N-1}(\gamma)\cdot \nabla^2 h({\bm\sigma}^0)\cdot {\bf e}_{N-1}({\bm\sigma}^0)
        \bigg].
    \end{split}
\end{align}
One can check explicitly that this expression vanishes at the point $\gamma$ where the geodesic energy profile reaches its maximum.

Consider now the orthogonal component of the gradient. In the notation of Appendix \ref{app:statistics_HESSIANS}, we see that component-wise in the chosen local basis it holds for $i<N-1$:
\begin{align}
    \frac{\nabla h({\bm\sigma}(\gamma))}{\sqrt{N-1}}\cdot{\bf x}_i=-\gamma\beta\,\sqrt{1-q^2}{\bf x}_i\cdot \frac{\nabla^2 h({\bm\sigma}^0)}{\sqrt{N-1}}\cdot {\bf e}_{N-1}({\bm\sigma}^0)=-\gamma\beta\sqrt{1-q^2}\,m_{i,N-1}^0.
\end{align}
Therefore, the orthogonal component ${\bf g}^\perp(\gamma)$ is proportional to the vector that makes up the last column of the shifted Hessian $\tilde{\mathcal{H}}({\bm \sigma}^0)$ (neglecting the last component of the column). This vector is in general not vanishing at the point that corresponds to the maximum of the energy profile along the geodesic. We define the normalized vector 
\begin{equation}
    {\bf v}_{\rm Hess}=  Z \sum_{i=1}^{N-2} [{\bf x}_i \cdot \tilde{\mathcal{H}}({\bm \sigma}^0) \cdot {\bf e}_{N-1}({\bm \sigma}^0)] \, {\bf x}_i =\frac{1}{\sqrt{\sum_{i=1}^{N-2} [m_{i \, N-1}^0]^2}} \tonde{m_{1 \, N-1}^0, m_{2 \, N-1}^0, \cdots, m_{N-2 \, N-1}^0,0}^T,
\end{equation}
which is orthogonal to both ${\bm \sigma}^a$ ($Z$ is the normalization factor). Plugging this into \eqref{eq:PrePAth} and making use of the fact that the entries $m_{i N-1}^a$ of the Hessians are uncorrelated to all other entries of the Hessian matrices, we see that the third line in Eq. \eqref{eq:PrePAth} vanishes on average, while in the second line we have:
\begin{equation}
    \mathbb{E} \quadre{ {\bf v}_{\rm Hess} \cdot \frac{\tilde{\mathcal{H}}({\bm \sigma}^0)}{\sqrt{2N}} \cdot {\bf e}_{N-1}({\bm \sigma}^0) }=  \mathbb{E} \quadre{\sqrt{\frac{\sum_{i=1}^{N-2} [m_{i \, N-1}^0]^2}{2 N}}}= \sqrt{\frac{\Delta^2}{2}}+\mathcal{O}\tonde{\frac{1}{N}}\stackrel{p=3}{=} \sqrt{\frac{3(1-q^2)}{1+q^2}}+\mathcal{O}\tonde{\frac{1}{N}},
\end{equation}
which leads to Eq. \ref{eq:energy_profile_Hess} in the main text. \\
Finally, we determine the overlap between $ {\bf v}_{\rm Hess}$ and  ${\bf v}_{\rm soft}^1$, to show that the two vectors are correlated when $u^1_{\rm typ} \neq 0$. It holds:
\begin{equation}
{\bf v}_{\rm Hess} \cdot {\bf v}_{\rm soft}^1= {\bf e}_{N-1}({\bm \sigma}^0) \cdot \frac{\nabla^2 h ({\bm \sigma}^0)}{\sqrt{N-1}} \cdot{\bf v}_{\rm soft}^1=\frac{q {\bm \sigma}^0-{\bm \sigma}^1}{\sqrt{1-q^2}} \cdot \frac{\nabla^2 h ({\bm \sigma}^0)}{\sqrt{N-1}} \cdot{\bf v}_{\rm soft}^1= - \frac{{\bm \sigma}^1}{\sqrt{1-q^2}} \cdot \frac{\nabla^2 h ({\bm \sigma}^0)}{\sqrt{N-1}} \cdot{\bf v}_{\rm soft}^1,
\end{equation}
which is equivalent to \eqref{eq:sotto} and thus of $\mathcal{O}(1)$ and non-vanishing when $u^1_{\rm typ} \neq 0$, as claimed in the main text. 

\end{document}